\documentclass[reprint,prd,twocolumn,superscriptaddress]{revtex4-1}
\pdfoutput=1
\usepackage{graphicx}
\newcommand{\Li}[1]{\mathop{\mathrm{Li}}\nolimits_{#1}}
\usepackage{times}
\usepackage{blindtext}
\begin{document}

\mbox{}\hfill DESY 20-092, P3H-20-023, SAGEX-20-14, TTP20-024

\title{Matching the heavy-quark fields in QCD and HQET at four loops}
\author{Andrey G.~Grozin}
\email{A.G.Grozin@inp.nsk.su}
\affiliation{Budker Institute of Nuclear Physics, Lavrentyev St.~11, Novosibirsk 630090, Russia}
\affiliation{Novosibirsk State University, Pirogov St.~1, Novosibirsk 630090, Russia}
\author{Peter Marquard}
\email{peter.marquard@desy.de}
\affiliation{Deutsches Elektronen-Synchrotron, DESY, 15738 Zeuthen, Germany}
\author{\\Alexander V.~Smirnov}
\email{asmirnov80@gmail.com}
\affiliation{Research Computing Center, Moscow State University, 119991 Moscow, Russia}
\author{Vladimir A.~Smirnov}
\email{smirnov@theory.sinp.msu.ru}
\affiliation{Skobeltsyn Institute of Nuclear Physics of Moscow State University, 119991 Moscow, Russia}
\author{Matthias Steinhauser}
\email{matthias.steinhauser@kit.edu}
\affiliation{Institut f\"ur Theoretische Teilchenphysik, Karlsruher Institut f\"ur Technologie (KIT), 76128 Karlsruhe, Germany}
%\date{}
\begin{abstract}
The QCD/HQET matching coefficient for the heavy-quark field
is calculated up to four loops.
It must be finite;
this requirement produces analytical results for some terms
in the four-loop on-shell heavy-quark field renormalization constant
which were previously only known numerically.
The effect of a non-zero lighter-flavor mass is calculated up to three loops.
A class of on-shell integrals with two masses is analyzed in detail.
By specifying our result to QED, we obtain the relation
between the electron field and the Bloch--Nordsieck field
with four-loop accuracy.
\end{abstract}
\maketitle

%- }}}
%- {{{ Introduction:

\section{Introduction}
\label{S:Intro}

Some classes of QCD problems with a single heavy quark can be examined
in a simpler effective theory, the so-called heavy quark effective theory
(HQET, see, e.\,g., \cite{Neubert:1993mb,Manohar:2000dt,Grozin:2004yc}).
Let us consider QCD with a single heavy flavor $Q$ and $n_l$ light flavors
($n_f = n_l + n_h$, $n_h = 1$).
The heavy-quark momentum can be decomposed as $p = Mv + k$,
where $M$ is the on-shell $Q$ mass, and $v$ is some reference 4-velocity ($v^2 = 1$).
In the case of QED, it is called Bloch--Nordsieck effective theory~\cite{Bloch:1937pw}.

In the effective theory, the heavy quark (respectively lepton) is represented by the field $h_v$.
The $\overline{\text{MS}}$ renormalized fields $Q(\mu)$ and $h_v(\mu)$ are related by~\cite{Grozin:2010wa}
\begin{equation}
Q(\mu) = e^{-iMvx} \left[ \sqrt{z(\mu)} \left( 1 + \frac{\rlap{\,/}D_\bot}{2 M} \right) h_v(\mu) + \mathcal{O}\left(\frac{1}{M^2}\right) \right]\,,
\label{Intro:Match}
\end{equation}
where $D_\bot^\mu = D^\mu - v^\mu\,v\cdot D$,
and the matching coefficient is given by
\begin{equation}
z(\mu) = \frac{Z_h(\alpha_s^{(n_l)}(\mu),\xi^{(n_l)}(\mu))\,Z_Q^{\text{os}}(g_0^{(n_f)},\xi_0^{(n_f)})}%
{Z_Q(\alpha_s^{(n_f)}(\mu),\xi^{(n_f)}(\mu))\,Z_h^{\text{os}}(g_0^{(n_l)},\xi_0^{(n_l)})}\,.
\label{Intro:z}
\end{equation}
Here $Z_Q^{\text{os}}$ and $Z_h^{\text{os}}$ are the on-shell field renormalization constants
(they depend on the corresponding bare couplings and bare gauge-fixing parameters),
and $Z_Q$ and $Z_h$ are the $\overline{\text{MS}}$ renormalization constants.
The covariant-gauge fixing parameter is defined in such a way that the bare gluon propagator is
given by $(g_{\mu\nu} - \xi_0 p_\mu p_\nu/p^2)/p^2$;
it is renormalized by the gluon-field renormalization constant:
$1-\xi_0 = Z_A(\alpha_s(\mu),\xi(\mu)) (1-\xi(\mu))$.
The $1/M$ correction in~(\ref{Intro:Match}) is fixed by reparametrization invariance~\cite{Luke:1992cs}.

The $\overline{\text{MS}}$ renormalized matching coefficient is obviously
finite at $\varepsilon \to 0$, because it relates the off-shell renormalized
propagators in the two theories, which are both finite.  The ultraviolet
divergences cancel in the ratios $Z_Q/Z_Q^{\text{os}}$ and
$Z_h/Z_h^{\text{os}}$, because they relate renormalized fields; the infrared
divergences cancel in $Z_Q^{\text{os}}/Z_h^{\text{os}}$, because HQET is
constructed to reproduce the infrared behavior of QCD;
the $\overline{\text{MS}}$ renormalization constants $Z_Q$ and $Z_h$
(purely off-shell quantities)
are infrared finite.  If we assume that all light flavors are massless we have
$Z_h^{\text{os}} = 1$: all loop corrections vanish because they contain no
scale, ultraviolet and infrared divergences of $Z_h^{\text{os}}$ mutually
cancel.  Taking light-quark masses $m_i$ into account produces corrections
suppressed by powers of $m_i/M$, see Sect.~\ref{S:Mass}.

The matching coefficient satisfies the renormalization-group equation
\begin{eqnarray}
&&\frac{d \log z(\mu)}{d \log \mu} =
\nonumber\\
&&\gamma_h(\alpha_s^{(n_l)}(\mu),\xi^{(n_l)}(\mu))
- \gamma_Q(\alpha_s^{(n_f)}(\mu),\xi^{(n_f)}(\mu))\,,
\label{Intro:RG}
\end{eqnarray}
where the anomalous dimensions are defined as
$\gamma_i = d \log Z_i / d \log \mu$ ($i = Q$, $h$).
It is sufficient to obtain the initial condition $z(\mu_0)$ for some scale $\mu_0 \sim M$;
$z(\mu)$ for other renormalization scales $\mu$ can be found by solving Eq.~(\ref{Intro:RG}).
We choose to present the result for $\mu_0=M$.

The heavy-quark field matching coefficient $z(\mu)$
has been calculated up to three loops~\cite{Grozin:2010wa}.
When the matching coefficient is used within a quantity containing $1/\varepsilon$ divergences,
terms with positive powers of $\varepsilon$ in $z(\mu)$ are needed;
such terms were not given in~\cite{Grozin:2010wa}.
We present the four-loop result in Sect.~\ref{S:QCD}.
Power corrections due to lighter-flavor masses up to three loops
are obtained in Sect.~\ref{S:Mass}.
The QED result, i.\,e.\ the four-loop relation between the lepton field and the Bloch--Nordsieck field,
is discussed in Sect.~\ref{S:QED}.
%This QED matching coefficient is gauge-invariant to all orders,
%as proved in~\cite{Grozin:2010wa}.
In Appendix~\ref{S:Dec} we provide analytic results for the decoupling
  coefficients for the strong coupling constant and the gluon field up to
  three-loop order including linear $\varepsilon$ terms.
Appendix~\ref{S:2m} contains a detailed analysis of a class of on-shell integrals with two masses.
It allows us, in particular, to obtain exact results for the three-loop
  term in the $\overline{\text{MS}}$--on-shell mass relation 
  with a closed massless and a closed lighter-flavor massive fermion loop
(previously this term was only known as a truncated series in this mass ratio).

%- }}}
%- {{{ The QCD and HQET heavy-quark fields:

\section{The QCD and HQET heavy-quark fields}
\label{S:QCD}

If we assume that all light flavors are massless, then~(\ref{Intro:z}) gives
\begin{eqnarray}
&&\log z(\mu) = \log Z_Q^{\text{os}}(g_0^{(n_f)},\xi_0^{(n_f)})
\label{QCD:z}\\
&&{} - \log Z_Q(\alpha_s^{(n_f)}(\mu),\xi^{(n_f)}(\mu))
+ \log Z_h(\alpha_s^{(n_l)}(\mu),\xi^{(n_l)}(\mu))\,.
\nonumber
\end{eqnarray}
The on-shell heavy-quark field renormalization constant $Z_Q^{\text{os}}$
depends on the bare coupling $g_0^{(n_f)}$, the bare gauge parameter $\xi_0^{(n_f)}$
and the on-shell mass $M$:
\begin{eqnarray}
&&Z_Q^{\text{os}} = 1 + \sum_{L=1}^\infty \left(4 \frac{\bigl(g_0^{(n_f)}\bigr)^2 M^{-2\varepsilon}}{(4\pi)^{d/2}} e^{-\gamma_E\varepsilon}\right)^L Z_L\,,
\nonumber\\
&&Z_L = \sum_{n=0}^\infty Z_{L,n}(\xi_0^{(n_f)}) \varepsilon^{n-L}\,.
\label{QCD:ZQ}
\end{eqnarray}
The two-loop expression is known exactly in $\varepsilon$~\cite{Broadhurst:1991fy};
it contains a single non-trivial master integral,
further terms of its $\varepsilon$ expansion are presented in~\cite{Broadhurst:1991fi,Broadhurst:1996az}.
The three-loop term has been calculated in~\cite{Melnikov:2000zc,Marquard:2007uj}.
At four loops, the terms with $n_l^3$ and $n_l^2$ are known analytically~\cite{Lee:2013sx},
and the remaining ones numerically~\cite{Marquard:2018rwx}.
Recently the QED-like color structures
$C_F^4$, $C_F^3 T_F n_h$, $C_F^2 (T_F n_h)^2$, $C_F^3 (T_F n_h)^3$, $d_{FF} n_h$
have been calculated analytically~\cite{Laporta:2020fog}.
Here and below we use the notation
\begin{equation}
d_{FF} = \frac{d_F^{abcd} d_F^{abcd}}{N_F}\,,\quad
d_{FA} = \frac{d_F^{abcd} d_A^{abcd}}{N_F}\,,
\label{QCD:dFF}
\end{equation}
where $N_R = \mathop{\mathrm{Tr}} \mathbf{1}_R$ (with $R=F$),
$d_R^{abcd} = \mathop{\mathrm{Tr}} t_R^{(a} t_R^{b\vphantom{(}}
t_R^{c\vphantom{)}} t_R^{d)}$
(with $R=F$ or $A$),
and the round brackets mean symmetrization
(for $SU(N_c)$ gauge group $d_{FF} = (N_c^2-1) (N_c^4-6N_c^2+18) / (96 N_c^3)$,
$d_{FA} = (N_c^2-1) (N_c^2+6) / 48$).
This result contains the same master integrals as the electron
  $g-2$~\cite{Laporta:2017okg,Marquard:2017iib}.
In~\cite{Laporta:2017okg}
they have been calculated numerically to 1100 digits,
and analytical expressions have been reconstructed using PSLQ.
In the case of the light-by-light contribution $d_{FF} n_h$
the results contain $\varepsilon^0$ terms of 6 master integrals
(known numerically to 1100 digits);
all the remaining constants are completely expressed via known transcendental numbers
(Note that the definition of the constant $t_{63}$ is missing
in the journal article~\cite{Laporta:2020fog};
it is included in the version v3 of the arXiv publication.).

The $\overline{\text{MS}}$ quark-field anomalous dimension $\gamma_q$ (and hence $\log Z_Q$)
is well known~\cite{Chetyrkin:1999pq,Czakon:2004bu,Luthe:2016xec,Baikov:2017ujl}.
The HQET field anomalous dimension $\gamma_h$ (and hence $\log Z_h$)
is known at three loops~\cite{Melnikov:2000zc,Chetyrkin:2003vi}.
At four loops, some color structures are known analytically:
$C_F (T_F n_l)^3$~\cite{Broadhurst:1994se},
$C_F^2 (T_F n_l)^2$~\cite{Grozin:2015kna,Grozin:2016ydd},
$C_F C_A (T_F n_l)^2$~\cite{Marquard:2018rwx},
$C_F^3 T_F n_l$~\cite{Grozin:2018vdn},
$d_F^{abcd} d_F^{abcd} n_l$~\cite{Grozin:2017css},
$C_F^2 C_A T_F n_l$ and $C_F C_A^2 T_F n_l$~\cite{Bruser:2019auj};
$C_F C_A^3$ and $d_F^{abcd} d_A^{abcd}$ are known numerically~\cite{Marquard:2018rwx}.

We need to express the three terms in~(\ref{QCD:z}) in terms of
the same set of variables, for which we choose $\alpha_s^{(n_f)}(\mu)$ and $\xi^{(n_f)}(\mu)$.
Expressing $g_0^{(n_f)}$ and $\xi_0^{(n_f)}$ via these variables is straightforward,
since the three-loop renormalization constants in QCD are well known.
Expressing $\alpha_s^{(n_l)}(\mu)$ and $\xi^{(n_l)}(\mu)$ via the $n_f$-flavor quantities
requires decoupling relations up to ${\cal O}(\varepsilon)$ at three loops.
For convenience we present explicit results in Appendix~\ref{S:Dec}.

The resulting matching coefficient $z(M)$ must be finite at $\varepsilon \to 0$.
This requirement together with the known results for $Z_Q$ and $Z_h$
leads to analytical expressions
for the four-loop coefficients $Z_{4,0}$, $Z_{4,1}$, and $Z_{4,2}$ in~(\ref{QCD:ZQ})
as well as for $Z_{4,3}$,
except two color structures $C_F C_A^3$ and $d_{FA}$
where the corresponding terms in $\gamma_h$ are not known analytically.
The analytic results are presented in the tables~\ref{T432} and~\ref{T1}.
We refrain from showing results for the $n_l^2$ and $n_l^3$ terms,
which are already known since a few years~\cite{Lee:2013sx}.
Furthermore, we have introduced $a_n = \Li{n}(1/2)$
(in particular $a_1 = \log 2$);
$\zeta_n$ denotes the Riemann zeta function and
$\xi_0 = \xi_0^{(n_f)}$.
Analytical results for the color structures
$C_F^4$, $C_F^3 T_F n_h$, $C_F^2 (T_F n_h)^2$, $C_F^3 (T_F n_h)^3$, $d_{FF} n_h$
were recently obtained~\cite{Laporta:2020fog}.
They agree with the expressions given in tables~\ref{T432} and~\ref{T1}.
Numerical results for these coefficients are given
in the tables~V, VI, and VII of Ref.~\cite{Marquard:2018rwx}.
Good agreement is found.

\begin{table*}
\caption{Coefficients $Z_{4,n}$ of the $1/\varepsilon^{4,3,2}$
    terms entering the four-loop result $Z_4$ in Eq.~(\ref{QCD:ZQ}).
    Note that the color structures $d_{FF} n_l$, $d_{FF} n_h$, $d_{FA}$ have zero coefficients.}
\label{T432}
\begin{ruledtabular}
\begin{tabular}{llll}
Color & $\varepsilon^{-4}$ & $\varepsilon^{-3}$ & $\varepsilon^{-2}$\\
\hline
$C_F^4$ & $\frac{27}{2048}$ & $\frac{171}{2048}$
& $\frac{3}{32} \left(3 \pi^2 a_1 - \frac{9}{2} \zeta_3 - \frac{153}{64} \pi^2 + \frac{1945}{256}\right)$\\
$C_F^3 C_A$ & $-\frac{99}{1024}$ & $-\frac{779}{1024}$
& $\frac{1}{64} \left(-119 \pi^2 a_1 + \frac{363}{2} \zeta_3 + \frac{1487}{16} \pi^2 - \frac{77405}{192}\right)$\\
$C_F^2 C_A^2$ & $\frac{1331}{6144}$ & $\frac{203}{96}$
& $\frac{1}{32} \left(\frac{649}{6} \pi^2 a_1 - \frac{2653}{16} \zeta_3 + \frac{\pi^4}{45} - \frac{46321}{576} \pi^2 + \frac{431051}{768}\right)
+ \frac{\xi_0}{128} \left(\frac{9}{8} \zeta_3 - \frac{\pi^4}{45} + \frac{1}{8}\right)$\\
$C_F C_A^3$ & $-\frac{1331}{9216}$ & $-\frac{97669}{55296}$
& $\frac{1}{32} \left(- \frac{121}{3} \pi^2 a_1 + \frac{7531}{128} \zeta_3 - \frac{127}{2160} \pi^4 + \frac{20449}{864} \pi^2 - \frac{419083}{864}\right)$\\
& & & ${} - \frac{59}{1024} \xi_0 \left(\frac{3}{8} \zeta_3 - \frac{\pi^4}{135} + \frac{1}{24}\right)
+ \frac{\xi_0^2}{1024} \left(\frac{9}{8}\zeta_3 - \frac{\pi^4}{45} + \frac{1}{8}\right)$\\
$C_F^3 T_F n_h$ & $\frac{9}{128}$ & $\frac{131}{512}$
& $\frac{1}{4} \left(3 \pi^2 a_1 - \frac{9}{2} \zeta_3 - \frac{87}{32} \pi^2 + \frac{1747}{192}\right)$\\
$C_F^2 C_A T_F n_h$ & $\frac{-67+\xi_0}{256}$ & $- \frac{1619 + \frac{29}{3} \xi_0}{1024}$
& $\frac{1}{8} \left( - \frac{53}{3} \pi^2 a_1 + \frac{433}{16} \zeta_3 + \frac{1697}{96} \pi^2 - \frac{257689}{2304}\right)
- \frac{\xi_0}{128} \left(3 \zeta_3 - \frac{\pi^2}{6} - \frac{137}{48}\right)$\\
$C_F C_A^2 T_F n_h$ & $\frac{441 - \frac{97}{9} \xi_0}{2048}$ & $\frac{\frac{216101}{36} + \frac{413}{12} \xi_0 + \xi_0^2}{3072}$
& $\frac{1}{12} \left(11 \pi^2 a_1 - \frac{2081}{128} \zeta_3 + \frac{11}{1080} \pi^4 - \frac{24725}{1536} \pi^2 + \frac{13306637}{55296}\right)$\\
& & & ${} + \frac{\xi_0}{16} \left(\frac{27}{64} \zeta_3 - \frac{\pi^4}{405} - \frac{97}{3456} \pi^2 - \frac{6251}{4608}\right)
- \frac{\xi_0^2}{6144} (\zeta_3 + 5)$\\
$C_F^2 (T_F n_h)^2$ & $\frac{3}{32}$ & $\frac{27}{128}$
& $\frac{\pi^2}{3} a_1 - \frac{\zeta_3}{2} - \frac{47}{96} \pi^2 + \frac{4337}{1536}$\\
$C_F C_A (T_F n_h)^2$ & $\frac{- 15 + \frac{\xi_0}{3}}{128}$ & $- \frac{\frac{1571}{18} + \xi_0}{128}$
& $\frac{1}{2} \left(- \frac{\pi^2}{3} a_1 + \frac{\zeta_3}{2} + \frac{779}{576} \pi^2 - \frac{14449}{864}\right)
+ \frac{\xi_0}{144} \left(\frac{\pi^2}{8} + \frac{17}{3}\right)$\\
$C_F (T_F n_h)^3$ & $\frac{1}{36}$ & $\frac{41}{864}$
& $\frac{1}{36} \left(-\frac{11}{3} \pi^2 + \frac{679}{16}\right)$\\
\hline
$C_F^3 T_F n_l$ & $\frac{9}{256}$ & $\frac{1}{4}$
& $\frac{1}{4} \left(\frac{5}{2} \pi^2 a_1 - 3 \zeta_3 - \frac{121}{64} \pi^2 + \frac{5491}{768}\right)$\\
$C_F^2 C_A T_F n_l$ & $-\frac{121}{768}$ & $-\frac{373}{256}$
& $\frac{1}{16} \left(-\frac{103}{3} \pi^2 a_1 + 41 \zeta_3 + \frac{3431}{144} \pi^2 - \frac{32869}{192}\right)$\\
$C_F C_A^2 T_F n_l$ & $\frac{121}{768}$ & $\frac{2903}{1536}$
& $\frac{1}{4} \left(\frac{11}{3} \pi^2 a_1 - \frac{859}{256} \zeta_3 + \frac{11}{4320} \pi^4 - \frac{715}{576}\pi^2 + \frac{410389}{6912}\right)
+ \frac{\xi_0}{64} \left(\frac{3}{8} \zeta_3 - \frac{\pi^4}{135} + \frac{1}{24}\right)$\\
$C_F^2 T_F^2 n_h n_l$ & $\frac{3}{32}$ & $\frac{65}{128}$
& $\frac{2}{3} \pi^2 a_1 - \frac{3}{4} \zeta_3 - \frac{21}{32} \pi^2 + \frac{2425}{576}$\\
$C_F C_A T_F^2 n_h n_l$ & $\frac{- 89 + \xi_0}{576}$ & $- \frac{\frac{4555}{18} + \xi_0}{192}$
& $ - \frac{\pi^2}{3} a_1 + \frac{\zeta_3}{4} + \frac{1063}{1728} \pi^2 - \frac{68323}{5184}
+ \frac{\xi_0}{288} \left(\frac{\pi^2}{6} + 11\right)$\\
$C_F T_F^3 n_h^2 n_l$ & $\frac{1}{24}$ & $\frac{19}{96}$
& $\frac{1}{72} \left(-13 \pi^2 + \frac{4895}{24}\right)$
\end{tabular}
\end{ruledtabular}
\end{table*}

\begin{table*}
\caption{Coefficients $Z_{4,3}$ of the $1/\varepsilon$ term
    entering the four-loop result $Z_4$ in Eq.~(\ref{QCD:ZQ}).
    Note that the color structures $C_F C_A^3$ and $d_{FA}$ are not known analytically.}
\label{T1}
\begin{ruledtabular}
\begin{tabular}{ll}
Color & $\varepsilon^{-1}$\\
\hline
$C_F^4$&
$\frac{1}{4} \left(57 a_4 + \frac{19}{8} a_1^4 - \frac{27}{4} \pi^2 a_1^2 - \frac{1571}{32} \pi^2 a_1 - \frac{25}{16} \zeta_5 - \frac{3}{8} \pi^2 \zeta_3 + \frac{5045}{128} \zeta_3 + \frac{101}{160} \pi^4 + \frac{33539}{1536} \pi^2 + \frac{23865}{2048}\right)$\\
$C_F^3 C_A$&
$- \frac{1}{4} \left(129 a_4 + \frac{43}{8} a_1^4 - \frac{467}{12} \pi^2 a_1^2 - \frac{33263}{192} \pi^2 a_1 + 25 \zeta_5 - \frac{157}{16} \pi^2 \zeta_3 + \frac{5477}{128} \zeta_3 + \frac{26707}{5760} \pi^4 + \frac{46967}{576} \pi^2 + \frac{245183}{3072}\right)$\\
$C_F^2 C_A^2$&
$- \frac{1}{16} \left(\frac{4171}{3} a_4 + \frac{4171}{72} a_1^4 + \frac{8257}{36} \pi^2 a_1^2 - \frac{20803}{144} \pi^2 a_1 - \frac{14733}{32} \zeta_5 + \frac{1739}{12} \pi^2 \zeta_3 + \frac{472475}{288} \zeta_3 - \frac{217663}{5760} \pi^4 - \frac{23849}{576} \pi^2 - \frac{2018473}{2304}\right)$\\
&${} - \frac{\xi_0}{64} \left(\frac{9}{16} \zeta_5 + \frac{\pi^2}{3} \zeta_3 - \frac{95}{32} \zeta_3 + \frac{383}{8640} \pi^4 - \frac{3}{16} \pi^2 - \frac{13}{12}\right)$\\
$C_F^3 T_F n_h$&
$a_4 + \frac{a_1^4}{24} - \frac{21}{8} \pi^2 a_1^2 - \frac{249}{16} \pi^2 a_1 + \frac{5}{16} \zeta_5 - \frac{\pi^2}{8} \zeta_3 + \frac{2705}{768} \zeta_3 + \frac{99}{320} \pi^4 + \frac{103157}{13824} \pi^2 + \frac{142385}{18432}$\\
$C_F^2 C_A T_F n_h$&
$\frac{278}{3} a_4 + \frac{139}{36} a_1^4 + \frac{761}{144} \pi^2 a_1^2 - \frac{469}{36} \pi^2 a_1 - \frac{545}{64} \zeta_5 + \frac{169}{64} \pi^2 \zeta_3 + \frac{661373}{9216} \zeta_3 - \frac{69871}{69120} \pi^4 + \frac{434299}{82944} \pi^2 - \frac{6467663}{110592}$\\
&${} - \frac{\xi_0}{2} \left(a_4 + \frac{a_1^4}{24} - \frac{\pi^2}{24} a_1^2 + \frac{59}{192} \zeta_3 - \frac{91}{11520} \pi^4 + \frac{29}{4608} \pi^2 - \frac{407}{6144}\right)$\\
$C_F C_A^2 T_F n_h$&
$- \frac{1}{3} \left(154 a_4 + \frac{77}{12} a_1^4 + \frac{515}{96} \pi^2 a_1^2 - \frac{335}{3} \pi^2 a_1 - \frac{4217}{384} \zeta_5 + \frac{1963}{576} \pi^2 \zeta_3 + \frac{1044785}{6144} \zeta_3 - \frac{283447}{276480} \pi^4 + \frac{857267}{12288} \pi^2 - \frac{498708329}{1327104}\right)$\\
&${} + \frac{\xi_0}{4} \left(a_4 + \frac{a_1^4}{24} - \frac{\pi^2}{24} a_1^2 - \frac{7}{144} \zeta_5 - \frac{\pi^2}{54} \zeta_3 + \frac{55}{3456} \zeta_3 - \frac{4631}{414720} \pi^4 + \frac{701}{27648} \pi^2 + \frac{59923}{36864}\right)
+ \frac{\xi_0^2}{3072} \left(\frac{7}{2} \zeta_3 - \frac{\pi^4}{120} + \frac{\pi^2}{3} + \frac{167}{8}\right)$\\
$C_F^2 (T_F n_h)^2$&
$- 20 a_4 - \frac{5}{6} a_1^4 - \frac{\pi^2}{6} a_1^2 + \frac{55}{36} \pi^2 a_1 - \frac{32131}{2304} \zeta_3 + \frac{53}{720} \pi^4 - \frac{11663}{51840} \pi^2 + \frac{167545}{13824}$\\
$C_F C_A (T_F n_h)^2$&
$12 a_4 + \frac{a_1^4}{2} - \frac{973}{72} \pi^2 a_1 + \frac{15}{16} \zeta_5 - \frac{11}{48} \pi^2 \zeta_3 + \frac{124621}{4608} \zeta_3 - \frac{\pi^4}{90} + \frac{1059347}{103680} \pi^2 - \frac{4538573}{82944}
- \frac{\xi_0}{24} \left(\frac{11}{24} \zeta_3 + \frac{\pi^2}{16} + \frac{29}{9}\right)$\\
$C_F (T_F n_h)^3$&
$\frac{1}{3} \left(2 \pi^2 a_1 - \frac{127}{9} \zeta_3 - \frac{7211}{4320} \pi^2 + \frac{71143}{3456}\right)$\\
$d_{FF} n_h$&
$- \frac{1}{8}$\\
\hline
$C_F^3 T_F n_l$&
$9 a_4 + \frac{3}{8} a_1^4 - \frac{37}{12} \pi^2 a_1^2 - \frac{1253}{96} \pi^2 a_1 + \frac{25}{32} \zeta_5 - \frac{\pi^2}{8} \zeta_3 + \frac{531}{128} \zeta_3 + \frac{1087}{2880} \pi^4 + \frac{6485}{1152} \pi^2 + \frac{2991}{1024}$\\
$C_F^2 C_A T_F n_l$&
$\frac{205}{3} a_4 + \frac{205}{72} a_1^4 + \frac{295}{36} \pi^2 a_1^2 - \frac{9289}{576} \pi^2 a_1 - \frac{605}{64} \zeta_5 + \frac{45}{16} \pi^2 \zeta_3 + \frac{75143}{1152} \zeta_3 - \frac{389}{270} \pi^4 + \frac{995}{192} \pi^2 - \frac{107579}{4608}$\\
$C_F C_A^2 T_F n_l$&
$- \frac{1}{4} \left(\frac{437}{3} a_4 + \frac{437}{72} a_1^4 + \frac{479}{36} \pi^2 a_1^2 - \frac{1631}{36} \pi^2 a_1 - \frac{6133}{256} \zeta_5 + \frac{3}{16} \zeta_3^2 + \frac{4057}{576} \pi^2 \zeta_3 + \frac{430895}{4608} \zeta_3 - \frac{854171}{414720} \pi^4 - \frac{26779}{3456} \pi^2 \right.$\\
&${} \left.{} - \frac{1583779}{5184} \right)  - \frac{\xi_0}{384} \left(\frac{19}{4} \zeta_5 + \frac{5}{3} \pi^2 \zeta_3 - \frac{29}{2} \zeta_3 + \frac{439}{2160} \pi^4 - \pi^2 - \frac{53}{12}\right)$\\
$C_F^2 T_F^2 n_h n_l$&
$- \frac{1}{3} \left(100 a_4 + \frac{25}{6} a_1^4 + \frac{23}{6} \pi^2 a_1^2 - \frac{209}{12} \pi^2 a_1 + \frac{6467}{96} \zeta_3 - \frac{8}{9} \pi^4 + \frac{26719}{3456} \pi^2 - \frac{184019}{4608}\right)$\\
$C_F C_A T_F^2 n_h n_l$&
$\frac{56}{3} a_4 + \frac{7}{9} a_1^4 + \frac{5}{9} \pi^2 a_1^2 - \frac{142}{9} \pi^2 a_1 + \frac{15}{16} \zeta_5 - \frac{11}{48} \pi^2 \zeta_3 + \frac{37781}{1728} \zeta_3 - \frac{47}{540} \pi^4 + \frac{105421}{10368} \pi^2 - \frac{9614047}{124416}
+ \frac{\xi_0}{24} \left(\frac{55}{36} \zeta_3 - \frac{\pi^2}{24} - 5\right)$\\
$C_F T_F^3 n_h^2 n_l$&
$\frac{1}{3} \left(4 \pi^2 a_1 - \frac{121}{6} \zeta_3 - \frac{1829}{480} \pi^2 + \frac{54391}{1152}\right)$\\
$d_{FF} n_l$&
$- \frac{1}{4} \left(\frac{5}{8} \zeta_5 - \frac{\pi^2}{3} \zeta_3 - \frac{\zeta_3}{2} + \frac{\pi^2}{3} + \frac{1}{2}\right)$
\end{tabular}
\end{ruledtabular}
\end{table*}

Using the matching coefficient $z(\mu)$ together with quantities
which contains $1/\varepsilon$ divergences,
terms with positive powers of $\varepsilon$ are needed.
In order to get the finite four-loop contribution,
we need the $\alpha_s^L$ term in $z(\mu)$ expanded up to $\varepsilon^{4-L}$.
Our result for $\mu=M$ is given by
\begin{widetext}
\begin{eqnarray}
&&z(M) = 1 - \frac{\alpha_s}{\pi} C_F
\biggl[1 + \varepsilon \biggl(\frac{\pi^2}{16} + 2\biggr)
    - \varepsilon^2 \biggl(\frac{\zeta_3}{4} - \frac{\pi^2}{12} - 4\biggr)
    - \varepsilon^3 \biggl(\frac{\zeta_3}{3} - \frac{3}{640} \pi^4 - \frac{\pi^2}{6} - 8\biggr)
    + \mathcal{O}(\varepsilon^4) \biggr]
\nonumber\\
&&{} + \left(\frac{\alpha_s}{\pi}\right)^2 C_F
\biggl\{
C_F \biggl(\pi^2 a_1 - \frac{3}{2} \zeta_3 - \frac{13}{16} \pi^2 + \frac{241}{128}\biggr)
- \frac{C_A}{2} \biggl(\pi^2 a_1 - \frac{3}{2} \zeta_3 - \frac{5}{8} \pi^2 + \frac{1705}{192}\biggr)
\nonumber\\
&&\qquad{}
- \frac{T_F n_h}{3} \biggl(\pi^2 - \frac{947}{96}\biggr)
+ \frac{T_F n_l}{12} \biggl(\pi^2 + \frac{113}{8}\biggr)
\nonumber\\
&&\quad{} + \varepsilon \biggl[
- C_F \biggl(24 a_4 + a_1^4 + 2 \pi^2 a_1^2 - \frac{23}{4} \pi^2 a_1 + \frac{147}{8} \zeta_3 - \frac{7}{20} \pi^4 + \frac{347}{128} \pi^2 + \frac{557}{256}\biggr)
\nonumber\\
&&\qquad{}
+ C_A \biggl(12 a_4 + \frac{a_1^4}{2} + \pi^2 a_1^2 - \frac{23}{8} \pi^2 a_1 + \frac{129}{16} \zeta_3 - \frac{7}{40} \pi^4 + \frac{769}{1152} \pi^2 - \frac{9907}{768}\biggr)
\nonumber\\
&&\qquad{}
+ T_F n_h \biggl(2 \pi^2 a_1 - 7 \zeta_3 - \frac{445}{288} \pi^2 + \frac{17971}{1728}\biggr)
+ T_F n_l \biggl(\zeta_3 + \frac{127}{288} \pi^2 + \frac{851}{192}\biggr)
\biggr]
\nonumber\\
&&\quad{} + \varepsilon^2 \biggl[
- C_F \biggl(144 a_5 + 138 a_4 - \frac{6}{5} a_1^5 + \frac{23}{4} a_1^4 - 4 \pi^2 a_1^3 + \frac{23}{2} \pi^2 a_1^2 + \frac{13}{15} \pi^4 a_1 - \frac{41}{2} \pi^2 a_1
\nonumber\\
&&\qquad\quad{}
- \frac{609}{4} \zeta_5 - \frac{11}{4} \pi^2 \zeta_3 + \frac{2061}{32} \zeta_3 - \frac{1555}{1536} \pi^4 + \frac{8947}{768} \pi^2 - \frac{1817}{512}\biggr)
\nonumber\\
&&\qquad{}
+ C_A \biggl(72 a_5 + 69 a_4 - \frac{3}{5} a_1^5 + \frac{23}{8} a_1^4 - 2 \pi^2 a_1^3 + \frac{23}{4} \pi^2 a_1^2 + \frac{13}{30} \pi^4 a_1 - \frac{41}{4} \pi^2 a_1
\nonumber\\
&&\qquad\quad{}
- \frac{609}{8} \zeta_5 - \frac{11}{8} \pi^2 \zeta_3 + \frac{7595}{288} \zeta_3 - \frac{14359}{23040} \pi^4 + \frac{6367}{2304} \pi^2 - \frac{79225}{1536}\biggr)
\nonumber\\
&&\qquad{}
- T_F n_h \biggl(48 a_4 + 2 a_1^4 + 4 \pi^2 a_1^2 - \frac{19}{2} \pi^2 a_1 + \frac{2405}{72} \zeta_3 - \frac{93}{320} \pi^4 + \frac{8605}{1728} \pi^2 - \frac{422747}{10368}\biggr)
\nonumber\\
&&\qquad{}
+ \frac{T_F n_l}{24} \biggl(\frac{305}{3} \zeta_3 + \frac{199}{80} \pi^4 + \frac{853}{24} \pi^2 + \frac{5753}{16}\biggr)
\biggr]
+ \mathcal{O}(\varepsilon^3)
\biggr\}
\nonumber\\
&&{} + \left(\frac{\alpha_s}{\pi}\right)^3 C_F
\biggl\{
- C_F^2 \biggl(28 a_4 + \frac{7}{6} a_1^4 - \frac{3}{2} \pi^2 a_1^2 - \frac{223}{12} \pi^2 a_1 + \frac{5}{16} \zeta_5 - \frac{\pi^2}{8} \zeta_3 + \frac{157}{8} \zeta_3 + \frac{19}{240} \pi^4 + \frac{4801}{576} \pi^2 + \frac{3023}{768}\biggr)
\nonumber\\
&&\qquad{}
- C_F C_A \biggl(\frac{a_4}{6} + \frac{a_1^4}{144} + \frac{181}{72} \pi^2 a_1^2 + \frac{43}{9} \pi^2 a_1 - \frac{145}{16} \zeta_5 + \frac{45}{16} \pi^2 \zeta_3 + \frac{289}{24} \zeta_3 - \frac{6697}{17280} \pi^4 - \frac{2137}{576} \pi^2 - \frac{24131}{4608}\biggr)
\nonumber\\
&&\qquad{} + \frac{C_A^2}{2} \biggl[
\frac{1}{3} \biggl(\frac{85}{2} a_4 + \frac{85}{48} a_1^4 + \frac{127}{24} \pi^2 a_1^2 - \frac{325}{24} \pi^2 a_1 - 37 \zeta_5 + \frac{127}{12} \pi^2 \zeta_3 + \frac{5857}{96} \zeta_3 - \frac{3419}{3840} \pi^4 - \frac{4339}{576} \pi^2 - \frac{1654711}{20736}\biggr)
\nonumber\\
&&\qquad\quad{}
+ \frac{\xi}{8} \biggl(\frac{7}{24} \zeta_5 + \frac{\pi^2}{9} \zeta_3 - \frac{13}{16} \zeta_3 + \frac{17}{1728} \pi^4 - \frac{\pi^2}{16} - \frac{13}{48}\biggr)
\biggr]
\nonumber\\
&&\qquad{}
+ C_F T_F n_h \biggl(12 a_4 + \frac{a_1^4}{2} - \frac{\pi^2}{2} a_1^2 + \frac{17}{9} \pi^2 a_1 + \frac{233}{288} \zeta_3 + \frac{31}{720} \pi^4 - \frac{553}{324} \pi^2 - \frac{13571}{3456}\biggr)
\nonumber\\
&&\qquad{}
- C_A T_F n_h \biggl[ \biggl(8 a_4 + \frac{a_1^4}{3} - \frac{\pi^2}{3} a_1^2 - \frac{80}{9} \pi^2 a_1 + \frac{15}{16} \zeta_5 - \frac{11}{48} \pi^2 \zeta_3 + \frac{2813}{576} \zeta_3 + \frac{17}{360} \pi^4 + \frac{9067}{1296} \pi^2 - \frac{788639}{41472}\biggr)
\nonumber\\
&&\qquad\quad{}
+ \frac{\xi}{24} \biggl(\zeta_3 - \frac{2387}{576}\biggr) \biggr]
\nonumber\\
&&\qquad{}
+ \frac{C_F T_F n_l}{3} \biggl(16 a_4 + \frac{2}{3} a_1^4 + \frac{4}{3} \pi^2 a_1^2 - \frac{47}{6} \pi^2 a_1 + \frac{137}{8} \zeta_3 - \frac{229}{720} \pi^4 + \frac{113}{24} \pi^2 + \frac{35}{6} \biggr)
\nonumber\\
&&\qquad{}
- \frac{C_A T_F n_l}{3} \biggl(8 a_4 + \frac{a_1^4}{3} + \frac{2}{3} \pi^2 a_1^2 - \frac{47}{12} \pi^2 a_1 + \frac{35}{24} \zeta_3 - \frac{19}{360} \pi^4 - \frac{13}{16} \pi^2 - \frac{111791}{5184}\biggr)
\nonumber\\
&&\qquad{}
+ \frac{(T_F n_h)^2}{3} \biggl(7 \zeta_3 + \frac{2}{15} \pi^2 - \frac{8425}{864}\biggr)
+ \frac{T_F^2 n_h n_l}{36} \biggl(13 \pi^2 - \frac{4721}{36}\biggr)
- \frac{(T_F n_l)^2}{18} \biggl(7 \zeta_3 + \frac{19}{6} \pi^2 + \frac{5767}{432}\biggr)
\nonumber\\
&&\quad{} + \varepsilon \biggl[
- C_F^2 \biggl(\frac{440}{3} a_5 - 16 \pi^2 a_4 + \frac{2444}{3} a_4 - \frac{11}{9} a_1^5 - \frac{2}{3} \pi^2 a_1^4 + \frac{611}{18} a_1^4 + \frac{115}{27} \pi^2 a_1^3 + \frac{2}{3} \pi^4 a_1^2 + \frac{2309}{36} \pi^2 a_1^2 - 14 \pi^2 \zeta_3 a_1
\nonumber\\
&&\qquad\quad{}
+ \frac{751}{432} \pi^4 a_1 - \frac{367}{2} \pi^2 a_1 - \frac{53}{2} \zeta_5 - \frac{29}{32} \zeta_3^2 - \frac{5861}{288} \pi^2 \zeta_3 + \frac{5119}{16} \zeta_3 + \frac{899}{5670} \pi^6 - \frac{54467}{34560} \pi^4 + \frac{74245}{2048} \pi^2 + \frac{19337}{1536}\biggr)
\nonumber\\
&&\qquad{}
- C_F C_A \biggl(\frac{487}{3} a_5 - 6 \pi^2 a_4 - \frac{1796}{9} a_4 - \frac{487}{360} a_1^5 - \frac{\pi^2}{4} a_1^4 - \frac{449}{54} a_1^4 - \frac{1135}{108} \pi^2 a_1^3 + \frac{\pi^4}{4} a_1^2 + \frac{7235}{216} \pi^2 a_1^2 - \frac{21}{4} \pi^2 \zeta_3 a_1
\nonumber\\
&&\qquad\quad{}
- \frac{949}{1080} \pi^4 a_1 + \frac{30803}{432} \pi^2 a_1 - \frac{125473}{384} \zeta_5 + \frac{143}{4} \zeta_3^2 + \frac{2703}{128} \pi^2 \zeta_3 - \frac{16339}{288} \zeta_3 + \frac{27331}{181440} \pi^6 - \frac{496741}{103680} \pi^4 - \frac{17665}{55296} \pi^2
\nonumber\\
&&\qquad\quad{}
- \frac{861659}{27648}\biggr)
\nonumber\\
&&\qquad{} + C_A^2 \biggl[
\frac{707}{6} a_5 - 7 \pi^2 a_4 + \frac{935}{9} a_4 - \frac{707}{720} a_1^5 - \frac{7}{24} \pi^2 a_1^4 + \frac{935}{216} a_1^4 - \frac{905}{216} \pi^2 a_1^3 + \frac{7}{24} \pi^4 a_1^2 + \frac{7081}{216} \pi^2 a_1^2 - \frac{49}{8} \pi^2 \zeta_3 a_1
\nonumber\\
&&\qquad\quad{}
- \frac{41}{8640} \pi^4 a_1 - \frac{8833}{864} \pi^2 a_1 - \frac{41569}{256} \zeta_5 + \frac{7451}{384} \zeta_3^2 + \frac{14915}{4608} \pi^2 \zeta_3 + \frac{67807}{3456} \zeta_3 + \frac{45047}{362880} \pi^6 - \frac{126391}{51840} \pi^4 - \frac{150229}{41472} \pi^2
\nonumber\\
&&\qquad\quad{}
- \frac{72476083}{746496}
- \frac{\xi}{128} \biggl(\frac{149}{6} \zeta_5 - \frac{25}{3} \zeta_3^2 - \frac{77}{72} \pi^2 \zeta_3 + \frac{63}{2} \zeta_3 - \frac{49}{405} \pi^6 - \frac{383}{1080} \pi^4 + \frac{35}{8} \pi^2 + \frac{35}{2}\biggr)
\biggr]
\nonumber\\
&&\qquad{}
+ C_F T_F n_h \biggl(72 a_5 - \frac{229}{6} a_4 - \frac{3}{5} a_1^5 - \frac{229}{144} a_1^4 + \pi^2 a_1^3 - \frac{2219}{144} \pi^2 a_1^2 + \frac{143}{180} \pi^4 a_1 + \frac{293}{6} \pi^2 a_1 - \frac{87}{8} \zeta_5 - \frac{81}{8} \pi^2 \zeta_3
\nonumber\\
&&\qquad\quad{}
- \frac{10913}{192} \zeta_3 + \frac{3649}{8640} \pi^4 - \frac{818609}{41472} \pi^2 + \frac{164069}{6912}\biggr)
\nonumber\\
&&\qquad{}
- C_A T_F n_h \biggl[48 a_5 - 8 \pi^2 a_4 + \frac{4247}{12} a_4 - \frac{2}{5} a_1^5 - \frac{\pi^2}{3} a_1^4 + \frac{4247}{288} a_1^4 + \frac{2}{3} \pi^2 a_1^3 + \frac{\pi^4}{3} a_1^2 + \frac{18133}{288} \pi^2 a_1^2 -7 \pi^2 \zeta_3 a_1
\nonumber\\
&&\qquad\quad{}
 + \frac{97}{180} \pi^4 a_1 - \frac{775}{9} \pi^2 a_1 + \frac{551}{64} \zeta_5 - \frac{181}{32} \zeta_3^2 - \frac{549}{64} \pi^2 \zeta_3 + \frac{88855}{384} \zeta_3 + \frac{1501}{15120} \pi^6 - \frac{12607}{5760} \pi^4 + \frac{286961}{13824} \pi^2
\nonumber\\
&&\qquad\quad{}
- \frac{35801821}{248832}
- \frac{\xi}{8} \biggl(\zeta_3 + \frac{\pi^4}{60} - \frac{121}{1728} \pi^2 - \frac{7367}{1152}\biggr)
\biggr]
\nonumber\\
&&\qquad{}
+ \frac{C_F T_F n_l}{3} \biggl(224 a_5 + \frac{1028}{3} a_4 - \frac{28}{15} a_1^5 + \frac{257}{18} a_1^4 - \frac{56}{9} \pi^2 a_1^3 + \frac{257}{9} \pi^2 a_1^2 - \frac{17}{90} \pi^4 a_1 - \frac{539}{9} \pi^2 a_1 - \frac{1027}{4} \zeta_5
\nonumber\\
&&\qquad\quad{}
- \frac{119}{16} \pi^2 \zeta_3 + \frac{1081}{6} \zeta_3 - \frac{18599}{8640} \pi^4 + \frac{160081}{4608} \pi^2 + \frac{3103}{72}\biggr)
\nonumber\\
&&\qquad{}
- C_A T_F n_l \biggl(\frac{112}{3} a_5 + \frac{514}{9} a_4 - \frac{14}{45} a_1^5 + \frac{257}{108} a_1^4 - \frac{28}{27} \pi^2 a_1^3 + \frac{257}{54} \pi^2 a_1^2 - \frac{17}{540} \pi^4 a_1 - \frac{539}{54} \pi^2 a_1 - \frac{859}{24} \zeta_5
\nonumber\\
&&\qquad\quad{}
- \frac{11}{16} \pi^2 \zeta_3 + \frac{1229}{432} \zeta_3 - \frac{3691}{6480} \pi^4 - \frac{1991}{648} \pi^2 - \frac{4500377}{93312}\biggr)
\nonumber\\
&&\qquad{}
+ \frac{(T_F n_h)^2}{3} \biggl(56 a_4 + \frac{7}{3} a_1^4 - \frac{7}{3} \pi^2 a_1^2 - \frac{4}{5} \pi^2 a_1 + \frac{3221}{80} \zeta_3 - \frac{31}{72} \pi^4 + \frac{39661}{7200} \pi^2 - \frac{636911}{8640}\biggr)
\nonumber\\
&&\qquad{}
+ T_F^2 n_h n_l \biggl(\frac{32}{3} a_4 + \frac{4}{9} a_1^4 + \frac{8}{9} \pi^2 a_1^2 - \frac{35}{9} \pi^2 a_1 + \frac{27}{2} \zeta_3 + \frac{179}{1080} \pi^4 + \frac{2245}{1296} \pi^2 - \frac{264817}{7776}\biggr)
\nonumber\\
&&\qquad{}
- \frac{(T_F n_l)^2}{54} \biggl(275 \zeta_3 + \frac{23}{5} \pi^4 + \frac{1081}{16} \pi^2 + \frac{253783}{864}\biggr)
\biggr]
+ \mathcal{O}(\varepsilon^2)
\biggr\}
\nonumber\\
&&{} + \left(\frac{\alpha_s}{\pi}\right)^4
\biggl\{
%- C_F^4 (3.86 \pm 1.5)
C_F^4 \biggl[L_0
-\frac{139}{2}a_5
+12\pi^2 a_4
-\frac{9137}{16}a_4
+\frac{139}{240}a_1^5
+\frac{\pi^2}{2}a_1^4
-\frac{9137}{384}a_1^4
-\frac{311}{72}\pi^2 a_1^3
-\frac{\pi^4}{2}a_1^2
-\frac{8597}{192}\pi^2 a_1^2
\nonumber\\
&&\qquad{}
+\frac{21}{2}\pi^2 \zeta_3 a_1
-\frac{2783}{2880}\pi^4 a_1
+\frac{33687}{256}\pi^2 a_1
-\frac{2937}{128}\zeta_5
+\frac{87}{128}\zeta_3^2
+\frac{2755}{192}\pi^2 \zeta_3
-\frac{113181}{512}\zeta_3
-\frac{899}{7560}\pi^6
+\frac{18553}{23040}\pi^4
\nonumber\\
&&\qquad{}
-\frac{24129}{1024}\pi^2
-\frac{90577}{8192}
\biggr]
\nonumber\\
&&\quad{}
+ C_F^3 C_A (14.12 \pm 3.6)
- C_F^2 C_A^2 \bigl[8.75607 \pm 2.9 - (0.00269 \pm 0.0012) \xi\bigr]
\nonumber\\
&&\quad{}
- C_F C_A^3 \bigl[142.552 \pm 0.82 - (0.43649 \pm 0.00076) \xi + (0.0205278 \pm 0.00012) \xi^2\bigr]
\nonumber\\
&&\quad{}
+ d_{FA} \bigl[9.4 \pm 2.1 + (0.147 \pm 0.013) \xi - (0.0748 \pm 0.0028) \xi^2\bigr]
\nonumber\\
&&\quad{}
%+ C_F^3 T_F (1.4632 \pm 0.019)
+ C_F^3 T_F n_h \biggl[L_1
+\frac{46}{3}a_5+16\pi^2 a_4
-\frac{35189}{48}a_4
-\frac{23}{180}a_1^5
+\frac{2}{3}\pi^2 a_1^4
-\frac{35189}{1152}a_1^4
-\frac{703}{108}\pi^2 a_1^3
-\frac{2}{3}\pi^4 a_1^2
-\frac{77155}{1152}\pi^2 a_1^2
\nonumber\\
&&\qquad{}
+14\pi^2 \zeta_3 a_1
-\frac{569}{2160}\pi^4 a_1
+\frac{3273}{16}\pi^2 a_1
-\frac{3067}{32}\zeta_5
+\frac{29}{32}\zeta_3^2
+\frac{2981}{288}\pi^2 \zeta_3
-\frac{119743}{384}\zeta_3
-\frac{899}{5670}\pi^6
+\frac{65953}{69120}\pi^4
\nonumber\\
&&\qquad{}
-\frac{572525}{13824}\pi^2
-\frac{305411}{36864}
\biggr]
\nonumber\\
&&\quad{}
+ C_F^2 C_A T_F n_h \bigl[14.893 \pm 0.083 - (0.657352 \pm 0.00024) \xi\bigr]
\nonumber\\
&&\quad{}
- C_F C_A^2 T_F n_h \bigl[3.1601 \pm 0.056 - (0.198984 \pm 0.00013) \xi + 0.0244254 \xi^2\bigr]
\nonumber\\
&&\quad{}
%+ C_F^2 T_F^2 (0.061555 \pm 0.00079)
+ C_F^2 (T_F n_h)^2 \biggl[L_2
+120a_5
+\frac{2749}{48}a_4
-a_1^5
+\frac{2749}{1152}a_1^4
-\frac{\pi^2}{3}a_1^3
-\frac{10525}{1152}\pi^2 a_1^2
+\frac{43}{36}\pi^4 a_1
+\frac{711}{20}\pi^2 a_1
-\frac{493}{8}\zeta_5
\nonumber\\
&&\qquad{}
-\frac{269}{24}\pi^2 \zeta_3
-\frac{10127}{2560}\zeta_3
-\frac{5513}{13824}\pi^4
-\frac{678719}{64800}\pi^2
-\frac{8452817}{414720}
\biggr]
\nonumber\\
&&\quad{}
- C_F C_A (T_F n_h)^2 \bigl[0.01995 \pm 0.0062 - 0.10436 \xi\bigr]
\nonumber\\
&&\quad{}
%+ 0.0064388 C_F T_F^3
+ C_F (T_F n_h)^3 \biggl[L_3
+ \frac{1}{3}\biggl(104a_4
+\frac{13}{3}a_1^4
+\frac{5}{3}\pi^2 a_1^2
-\frac{103}{10}\pi^2 a_1
+\frac{5881}{80}\zeta_3
-\frac{299}{360}\pi^4
+\frac{31451}{2700}\pi^2
-\frac{5981281}{51840}\biggr)
\biggr]
\nonumber\\
&&\quad{}
%+ d_{FF} (0.1 \pm 0.5)
+ d_{FF} n_h L_l
- C_F^3 T_F n_l (4.92605 \pm 0.0067)
+ C_F^2 C_A T_F n_l (15.0599 \pm 0.012)
\nonumber\\
&&\quad{}
+ C_F C_A^2 T_F n_l \bigl[166.421 \pm 0.031 - 0.134051 \xi\bigr]
- C_F^2 T_F^2 n_h n_l (5.08715 \pm 0.000074)
\nonumber\\
&&\quad{}
+ C_F C_A T_F^2 n_h n_l \bigl[0.53235 \pm 0.0015 + 0.0910988 \xi\bigr]
+ 0.0138079 C_F T_F^3 n_h^2 n_l
- d_{FF} n_l (2.18 \pm 0.8)
\nonumber\\
&&\quad{}
- C_F^2 (T_F n_l)^2 \biggl(\frac{32}{3} a_5 + \frac{188}{9} a_4 - \frac{4}{45} a_1^5 + \frac{47}{54} a_1^4 - \frac{8}{27} \pi^2 a_1^3 + \frac{47}{27} \pi^2 a_1^2 - \frac{31}{270} \pi^4 a_1 - \frac{239}{54} \pi^2 a_1 - \frac{601}{48} \zeta_5 - \frac{\pi^2}{2} \zeta_3
\nonumber\\
&&\qquad{}
+ \frac{6925}{576} \zeta_3 - \frac{1181}{10368} \pi^4 + \frac{1043}{384} \pi^2 + \frac{3146969}{497664}\biggr)
\nonumber\\
&&\quad{}
+ \frac{C_F C_A (T_F n_l)^2}{3} \biggl(16 a_5 + \frac{94}{3} a_4 - \frac{2}{15} a_1^5 + \frac{47}{36} a_1^4 - \frac{4}{9} \pi^2 a_1^3 + \frac{47}{18} \pi^2 a_1^2 - \frac{31}{180} \pi^4 a_1 - \frac{239}{36} \pi^2 a_1 - \frac{365}{32} \zeta_5 - \frac{11}{12} \pi^2 \zeta_3
\nonumber\\
&&\qquad{}
- \frac{1111}{64} \zeta_3 - \frac{4333}{17280} \pi^4 - \frac{6815}{1152} \pi^2 - \frac{4767085}{165888}\biggr)
\nonumber\\
&&\quad{}
+ \frac{C_F T_F^3 n_h n_l^2}{3} \biggl(\frac{\zeta_3}{16} - \frac{4}{15} \pi^4 + \frac{19}{27} \pi^2 + \frac{399325}{20736}\biggr)
+ \frac{C_F (T_F n_l)^3}{216} \biggl(\frac{467}{2} \zeta_3 + \frac{71}{20} \pi^4 + \frac{167}{3} \pi^2 + \frac{103933}{864}\biggr)
+ \mathcal{O}(\varepsilon)
\biggr\}
\nonumber\\
&&{}
+ \mathcal{O}(\alpha_s^5)\,,
\label{QCD:result}
\end{eqnarray}
\end{widetext}
where $\alpha_s = \alpha_s^{(n_f)}(M)$, $\xi = \xi^{(n_f)}(M)$.
$L_{0,l,1,2,3}$ are the $\varepsilon^0$ parts of the quantities 
$Z_2^{(4,0)}$, $Z_2^{(4,l)}$, $Z_2^{(4,1)}$, $Z_2^{(4,2)}$, $Z_2^{(4,3)}$
given in Eqs.~(28--32) of ~\cite{Laporta:2020fog}.
Their numerical values are given in Eqs.~(5--9) of that paper.
The finite four-loop terms of Eq.~(\ref{QCD:result}) are equal to
the corresponding finite four-loop terms in $Z_Q^{\text{os}}$
plus products of lower-loop quantities which are all known analytically.
For 14 out of 23 color structures these coefficients in $Z^{\text{os}}_Q$
are only known numerically~\cite{Marquard:2018rwx}.
We use these numerical values, together with their uncertainty estimates,
from the tables~V, VI, and VII of that paper.
Note that in Ref.~\cite{Marquard:2018rwx} $Z_Q^{\text{os}}$
has been computed in an expansion in $\xi$ up to the second order;
9 out of these 19 color structures are obviously gauge invariant,
and 7 more seem to be either gauge-invariant or have at most linear $\xi$ terms
(though we know no explicit proof).
The remaining 3 structures ($C_F C_A^3$, $d_{FA}$, $C_F C_A^2 T_F n_h$)
may contain terms with higher powers of $\xi$, which are not known.
The same is true for the corresponding terms in $z(\mu)$ in Eq.~(\ref{QCD:result}).

If we re-express $z(M)$ in Eq.~(\ref{QCD:result})  via $\alpha_s^{(n_l)}(M)$,
the terms up to three loops agree with~\cite{Grozin:2010wa}.
(Note that positive powers of $\varepsilon$
are not presented~\cite{Grozin:2010wa}.)
The $\alpha_s^4 n_l^3$ term also agrees with~\cite{Grozin:2010wa}.

After specifying the color factors to QCD with $N_c=3$
we obtain for $\varepsilon=0$
\begin{eqnarray}
&&z(M) = 1 - \frac{4}{3} \frac{\alpha_s}{\pi}
- \left(\frac{\alpha_s}{\pi}\right)^2
(17.45 - 1.33 n_l)
\nonumber\\
&&{} - \left(\frac{\alpha_s}{\pi}\right)^3
(262.42 - 0.78 \xi
- 35.81 n_l + 0.98 n_l^2)
\nonumber\\
&&{} - \left(\frac{\alpha_s}{\pi}\right)^4
\bigl[5137.52 - 15.67 \xi
+ 1.07 \xi^2
\nonumber\\
&&\quad{}
- \bigl(1030.82 - 0.71 \xi\bigr) n_l
+ 60.30 n_l^2 - 1.00 n_l^3 \bigr]
\nonumber\\
&&{} + \mathcal{O}(\alpha_s^5)\,.
\label{QCD:num}
\end{eqnarray}
In Landau gauge ($\xi^{(n_f)}=1$) at $n_l=4$ this gives
\begin{eqnarray}
&&z(M) = 1 - \frac{4}{3} \frac{\alpha_s}{\pi}
- 12.12 \left(\frac{\alpha_s}{\pi}\right)^2
- 134.11 \left(\frac{\alpha_s}{\pi}\right)^3
\nonumber\\
&&{} - 1903.22 \left(\frac{\alpha_s}{\pi}\right)^4
+ \mathcal{O}(\alpha_s^5)\,,
\label{QCD:num4}
\end{eqnarray}
while the naive nonabelianization~\cite{Broadhurst:1994se}
(large $\beta_0$ limit) predicts~\cite{Grozin:2010wa}
\begin{eqnarray}
&&1 - \frac{4}{3} \frac{\alpha_s}{\pi}
- 16.66 \left(\frac{\alpha_s}{\pi}\right)^2
- 153.41 \left(\frac{\alpha_s}{\pi}\right)^3
\nonumber\\
&&{} - 1953.40 \left(\frac{\alpha_s}{\pi}\right)^4
+ \mathcal{O}(\alpha_s^5)\,.
\label{QCD:nna}
\end{eqnarray}
The comparison to Eq.~(\ref{QCD:num4}) shows that up to four loops these predictions are rather good.
The coefficients are all negative and grow very fast,
which can be explained by the infrared renormalon at $u=1/2$~\cite{Grozin:2010wa}.
This is the closest possible position of a renormalon singularity in the Borel plane $u$ to the origin,
and it leads to the fastest possible growth of perturbative terms $(L-1)!\,(\beta_0/2)^L (\alpha_s/\pi)^L$.
The coefficients of powers of $\xi$ are much smaller
than the $\xi$-independent terms.

%- }}}
%- {{{ Effect of a lighter-flavor mass:

\section{Effect of a lighter-flavor mass}
\label{S:Mass}

Now we suppose that $n_m$ light flavors have a non-zero mass $m$,
while the remaining $n_0 = n_l-n_m$ light flavors are massless.
In practice, $n_m = 1$, e.\,g.\ $c$ in $b$-quark HQET.
In this case the massless result~(\ref{QCD:result}) for the matching coefficient
should be multiplied by the additional factor
\begin{eqnarray}
z' &=& \frac{Z^{\text{os}}_Q(g_0^{(n_f)},\xi_0^{(n_f)},m_0^{(n_f)})}{Z^{\text{os}}_Q(g_0^{(n_f)},\xi_0^{(n_f)},0)}
\nonumber\\
&&{}\times\frac{Z^{\text{os}}_h(g_0^{(n_l)},\xi_0^{(n_l)},0)}{Z^{\text{os}}_h(g_0^{(n_l)},\xi_0^{(n_l)},m_0^{(n_l)})}\,,
\label{Mass:z1}
\end{eqnarray}
where $Z^{\rm os}_{Q,h}(\ldots,0)\equiv Z^{\rm os}_{Q,h}(\ldots)$ in
Eq.~(\ref{Intro:z}) and $Z^{\text{os}}_h(g_0^{(n_l)},\xi_0^{(n_l)},0) = 1$.
This factor does not depend on the renormalization scale $\mu$.
In the expression
\begin{eqnarray}
&&\log z'
= \log Z^{\text{os}}_Q(g_0^{(n_f)},\xi_0^{(n_f)},m_0^{(n_f)})
\label{Mass:log}\\
&&{} - \log Z^{\text{os}}_Q(g_0^{(n_f)},\xi_0^{(n_f)},0)
- \log Z^{\text{os}}_h(g_0^{(n_l)},\xi_0^{(n_l)},m_0^{(n_l)})
\nonumber
\end{eqnarray}
we re-express all terms via $\alpha_s^{(n_f)}(M)$, $\xi^{(n_f)}(M)$
and the on-shell lighter-flavor mass $m$
(it is the same in both $n_f$ and $n_l$ flavor theories).
The result depends on the dimensionless ratio
\begin{equation}
x = \frac{m}{M}\,.
\label{Mass:x}
\end{equation}
If we express $z'$ via $\alpha_s^{(n_f)}(\mu)$, $\xi^{(n_f)}(\mu)$,
the coefficients will depend on $\mu$.
This dependence is determined by the renormalization-group equation
\begin{equation}
\frac{d \log z'}{d \log \mu} = 0
\label{Mass:RG}
\end{equation}
together with
\begin{eqnarray*}
&&\frac{d \log \alpha_s^{(n_f)}(\mu)}{d \log \mu} =
- 2 \varepsilon - 2 \beta^{(n_f)}(\alpha_s^{(n_f)}(\mu))\,,\\
&&\frac{d \log(1 - \xi^{(n_f)}(\mu))}{d \log \mu} =
- \gamma_A^{(n_f)}(\alpha_s^{(n_f)}(\mu),\xi^{(n_f)}(\mu))\,.
\end{eqnarray*}

\begin{figure}[h]
\begin{picture}(70,24)
\put(15,14){\makebox(0,0){\includegraphics{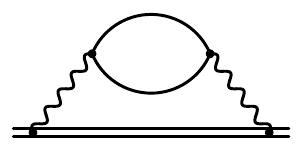}}}
\put(15,22){\makebox(0,0){$m$}}
\put(15,0){\makebox(0,0)[b]{a}}
\put(55,14){\makebox(0,0){\includegraphics{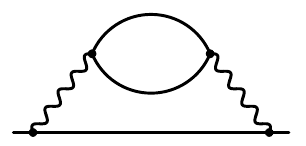}}}
\put(55,22){\makebox(0,0){$m$}}
\put(55,6){\makebox(0,0){$M$}}
\put(55,0){\makebox(0,0)[b]{b}}
\end{picture}
\caption{Two-loop contributions to the on-shell wave-function renormalization constants:
(a) in HQET; (b) in QCD.}
\label{F:disc}
\end{figure}
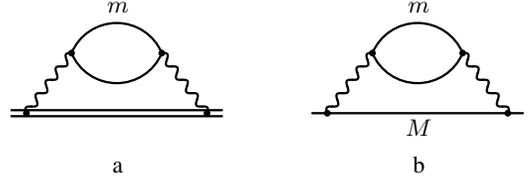

Ultraviolet divergences cancel in each fraction in~(\ref{Mass:z1}).  On the
other hand, the on-shell wave-function renormalization factors have extra
infrared divergences at $m=0$.  However, $z^\prime$ in
  Eq.~(\ref{Mass:z1}) has a smooth limit for $x\to0$. In the following we
  illustrate the cancellation for infrared divergences at two-loop
  order. Similar mechanisms are also at work at higher loop orders.  For
  dimensional reasons the two-loop corrections in Fig.~\ref{F:disc}a lead to
  $\log Z_h^{\text{os}}(m) \sim g_0^4 m^{-4\varepsilon}$.  Furthermore, we
  have $\log Z_h^{\text{os}}(0) = 0$. Thus, the limit $x\to0$ is
discontinuous.  In QCD (Fig.~\ref{F:disc}b) we have
$\log Z_Q^{\text{os}}(0) \sim g_0^4 M^{-4\varepsilon}$ for dimensional
  reasons.  For $m\ll M$ there are 3 regions (see~\cite{Beneke:1997zp,Smirnov:2002pj}):
\begin{itemize}
\item Hard (all momenta $\sim M$): a regular series in $m^2$,
$\log Z_Q^{\text{os}}(m)\big|_{\text{hard}} = \log Z_Q^{\text{os}}(0) \left[1 + \mathcal{O}(x^2)\right]$.
\item Soft-hard (momentum of one $m$-line is $\sim m$, all the remaining momenta are $\sim M$).
If we take the term $m$ from the numerator $\rlap/k+m$ of the soft propagator,
there is another factor $m$ in the numerator of the hard mass-$m$ propagator,
and the soft-loop integral is $\sim m^{2-2\varepsilon}$;
if we take $\rlap/k$ instead, we have to expand the hard subdiagram in $k$
up to the linear term, and the soft loop is $\sim m^{4-2\varepsilon}$.
We obtain
$\log Z_Q^{\text{os}}(m)\big|_{\text{soft-hard}} \sim g_0^4 M^{-2\varepsilon} m^{-2\varepsilon} x^4$.
\item Soft (all momenta $\sim m$): the leading term is the HQET one,
the Taylor series is in $x$ (not in $x^2$),
$\log Z_Q^{\text{os}}(m)\big|_{\text{soft}} = \log Z_h^{\text{os}}(m) \left[1 + \mathcal{O}(x)\right]$.
\end{itemize}
As a result, $\log Z_Q^{\text{os}}(m)\big|_{\text{hard}} - \log Z_Q^{\text{os}}(0)$ is smooth at $x\to0$;
$\log Z_Q^{\text{os}}(m)\big|_{\text{soft-hard}}$ is subleading and hence smooth;
$\log Z_Q^{\text{os}}(m)\big|_{\text{soft}}$ has the same discontinuity as $\log Z_h^{\text{os}}$;
hence $\log z'$~(\ref{Mass:log}) has a smooth limit 1 at $x\to0$.

The two-loop term in $Z^{\text{os}}_Q(g_0^{(n_f)},\xi_0^{(n_f)},m_0^{(n_f)})$
has been calculated up to $\varepsilon^0$ in~\cite{Broadhurst:1991fy}; the
result exact in $\varepsilon$ has been obtained in~\cite{Davydychev:1998si}.
The three-loop term has been calculated up to $\varepsilon^0$
in~\cite{Bekavac:2007tk}.  Some master integrals are only known as truncated
series in $x$ or as numerical interpolations, see~\cite{Bekavac:2009gz} for
detailed discussion of these master integrals.  Exact results in $x$ for
  the coefficient of $C_F T_F^2 n_m n_0 \alpha_s^3$ can be obtained using the
formulas of Appendix~\ref{S:2m}.

The HQET renormalization constant
$Z^{\text{os}}_h(g_0^{(n_l)},\xi_0^{(n_l)},m_0^{(n_l)})$ at two loops has been
calculated in~\cite{Broadhurst:1994se}, and at three loops
in~\cite{Grozin:2006xm} (one of the master integrals is discussed
in~\cite{Grozin:2007ap}; note that there are some typos in formulas in the
journal version of~\cite{Grozin:2006xm} fixed later in arXiv).

\begin{figure}[ht]
\begin{picture}(60,30)
\put(25,15){\makebox(0,0){\includegraphics[width=5cm]{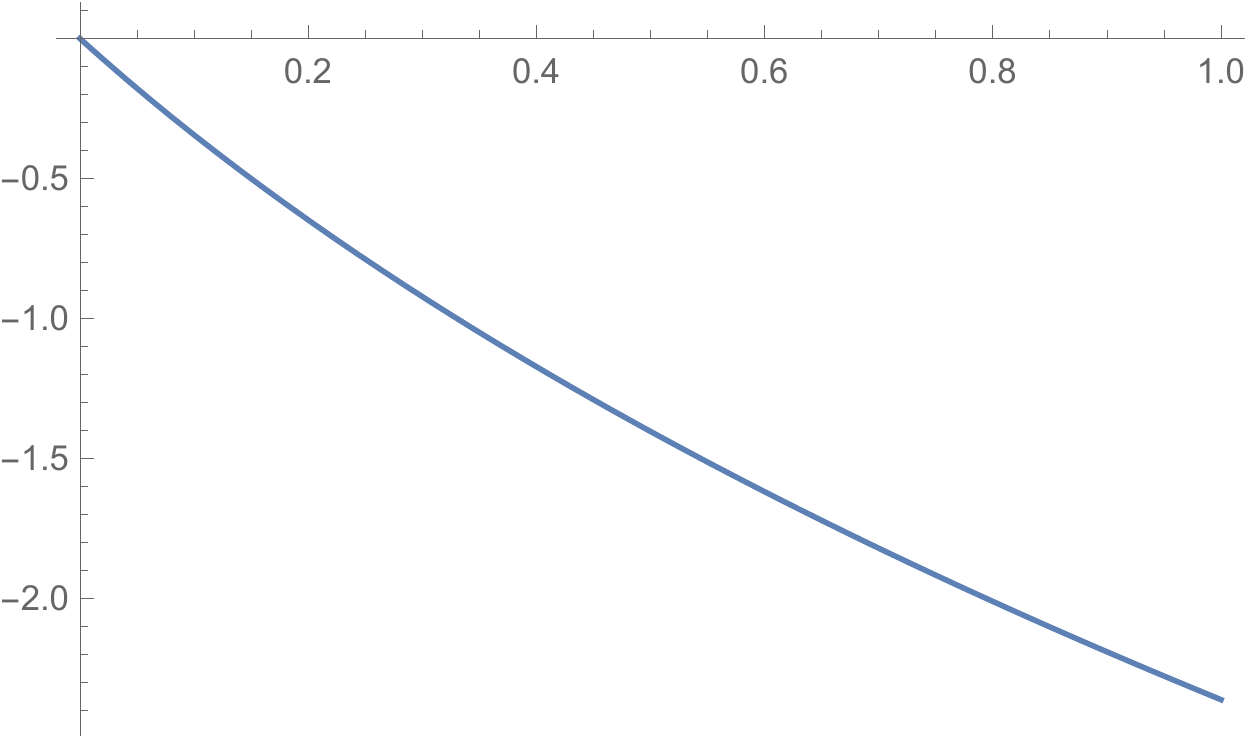}}}
\put(55,27){\makebox(0,0){$x$}}
\end{picture}
\caption{The function $A_0(x)$.}
\label{F:2l}
\end{figure}

Altogether we are now in the position to obtain $z'$ up to three
loops. The expansion of $z^\prime$ in terms of $\alpha_s^{(n_f)}(M)$ and
its decomposition into color factors is given by
\begin{widetext}
\begin{eqnarray}
z' &=& 1 + C_F T_F \left(\frac{\alpha_s^{(n_f)}(M)}{\pi}\right)^2
\left(A_0 + A_1 \varepsilon + \mathcal{O}(\varepsilon^2)\right)
\nonumber\\
&&{} + C_F T_F \left(\frac{\alpha_s^{(n_f)}(M)}{\pi}\right)^3
\bigl(C_F A_F + C_A A_A
+ T_F n_0 A_l + T_F n_m A_m + T_F n_h A_h
+ \mathcal{O}(\varepsilon)\bigr)
+ \mathcal{O}(\alpha_s^4)\,,
\label{Mass:result}
\end{eqnarray}
where
\begin{eqnarray}
A_0 &=& \frac{1}{4} \biggl[
(1-x) (2-x-x^2-6x^3) H_{1,0}(x)
- (1+x) (2+x-x^2+6x^3) H_{-1,0}(x)
\nonumber\\
&&{} - \frac{3}{2} \pi^2 x
+ (4 \log x + 7) x^2
- \frac{5}{2} \pi^2 x^3
+ (6 \log^2 x + \pi^2) x^4
\biggr]\,.
\label{Mass:A0}
\end{eqnarray}
The expansion of this function in $x$ reads
\begin{eqnarray}
A_0 &=& \frac{1}{4} \biggl[
- \frac{3}{2} \pi^2 x + 12 x^2 - \frac{5}{2} \pi^2 x^3
+ \left(6 \log^2 x - 11 \log x + \pi^2 + \frac{125}{12}\right) x^4
+ \sum_{n=3}^\infty \left(2 g(2n) \log x + \frac{d\,g(2n)}{d\,n}\right) x^{2n}
\biggr]\,,
\nonumber\\
g(x) &=& \frac{2}{x} - \frac{3}{x-1} - \frac{5}{x-3} + \frac{6}{x-4}\,.
\label{Mass:A0x}
\end{eqnarray}
Note that the only terms with odd powers of $x$ are $x^1$ and $x^3$.
The expansion in $x^{-1}$ is given by
\begin{equation}
A_0 = \frac{1}{4} \biggl[
- 2 \log^2 x^{-1} + \frac{19}{3} \log x^{-1} - \frac{\pi^2}{3} - \frac{229}{36}
+ \sum_{n=1}^\infty \left(2 g(-2n) \log x^{-1} + \frac{d\,g(-2n)}{d\,n}\right) x^{-2n}
\biggr]\,.
\label{Mass:A0i}
\end{equation}
For illustration we show in Fig.~\ref{F:2l} $A_0(x)$ for $x\in[0,1]$.
The ${\cal O}(\varepsilon)$ term at two loops reads
\begin{eqnarray}
A_1 &=& \frac{1}{4} \biggl[
(1-x) (2-x-x^2-6x^3) \bigl(2 H_{1,1,0}(x) - 4 H_{1,-1,0}(x)\bigr)
\nonumber\\
&&{} + (1+x) (2+x-x^2+6x^3) \bigl(2 H_{-1,-1,0}(x) - 4 H_{-1,1,0}(x) - \pi^2 H_{-1}(x)\bigr)
\nonumber\\
&&{} + (1-x) (9-6x+6x^2-17x^3) H_{1,0}(x)
- (1+x) (9+6x+6x^2+17x^3) H_{-1,0}(x)
\nonumber\\
&&{} + 4 x (3+5x^2) \bigl(H_{0,1,0}(x) + H_{0,-1,0}(x)\bigr)
+ 6 \pi^2 \left(L + 2 a_1 - \frac{5}{4}\right) x
+ \left(L + 2 \pi^2 + \frac{53}{2}\right) x^2
\nonumber\\
&&{} + 10 \pi^2 \left(L + 2 a_1 - \frac{23}{20}\right) x^3
- 12 \left(L^3 - \frac{17}{12} L^2 - \zeta_3 - \frac{17}{72} \pi^2\right) x^4
\biggr]
\nonumber\\
&=& \pi^2 \biggl(\frac{3}{2} L + 3 a_1 - \frac{19}{8}\biggr) x
+ \frac{5}{2} x^2
+ \pi^2 \biggl(\frac{5}{2} L + 5 a_1 - \frac{8}{3}\biggr) x^3
- \biggl(3 L^3 - \frac{17}{4} L^2 - \frac{3}{8} L - 3 \zeta_3 + \frac{2}{3} \pi^2 + \frac{2827}{288}\biggr) x^4
\nonumber\\
&&{} - \frac{63}{80} \pi^2 x^5
- \frac{2}{15} \biggl(\frac{61}{5} L - 2 \pi^2 - \frac{4243}{225}\biggr) x^6
- \frac{15}{112} \pi^2 x^7
- \frac{3}{56} \biggl(\frac{53}{35} L - \frac{3}{2} \pi^2 - \frac{5909}{1960}\biggr) x^8
+ \mathcal{O}(x^9)\,,
\label{Mass:A1}
\end{eqnarray}
where $L = \log x$.

At three-loop order the $C_F T_F^2 n_m n_0 \alpha_s^3$ term is known exactly via harmonic polylogarithms of $x$:
\begin{eqnarray}
A_l &=& \frac{1}{3} \biggl[
(1-x) (2-x-x^2-6 x^3) \biggl(H_{1,-1,0}(x) + \frac{\pi^2}{12} H_{1}(x)\biggr)
+ (1+x) (2+x-x^2+6 x^3) \biggl(H_{-1,1,0}(x) + \frac{5}{12} \pi^2 H_{-1}(x)\biggr)
\nonumber\\
&&{} - \frac{1}{6} (1-x) (19-11 x+x^2-39 x^3) H_{1,0}(x)
+ \frac{1}{6} (1+x) (19+11 x+x^2+39 x^3) H_{-1,0}(x)
\nonumber\\
&&{} - x (3+5 x^2) \left(H_{0,1,0}(x) + H_{0,-1,0}(x)\right)
- \pi^2 \biggl(\frac{3}{2} L + 3 a_1 - \frac{5}{2}\biggr) x
- \biggl(\frac{17}{2} L + 2 \pi^2 + \frac{91}{4}\biggr) \frac{x^2}{3}
- 5 \pi^2 \biggl(\frac{L}{2} + a_1 - \frac{2}{3}\biggr) x^3
\nonumber\\
&&{} + \biggl(2 L^3 - \frac{13}{2} L^2 - \pi^2 L - 9 \zeta_3 - \frac{13}{12} \pi^2\biggr) x^4
\biggr]
\nonumber\\
&=& - \pi^2 \biggl(\frac{L}{2} + a_1 - \frac{7}{6}\biggr) x
- \frac{7}{3} x^2
- \frac{5}{3} \pi^2 \biggl(\frac{L}{2} + a_1 - \frac{7}{12}\biggr) x^3
+ \biggl[\frac{2}{3} L^3 - \frac{13}{6} L^2 - \biggl(\frac{\pi^2}{3} - \frac{3}{4}\biggr) L
- 3 \zeta_3 + \frac{\pi^2}{4} + \frac{1175}{432}\biggr] x^4
\nonumber\\
&&{} + \frac{21}{40} \pi^2 x^5
+ \frac{4}{45} \biggl(\frac{13}{5} L - \frac{4}{3} \pi^2 - \frac{2414}{225}\biggr) x^6
+ \frac{5}{56} \pi^2 x^7
+ \biggl(\frac{4}{35} L - \frac{\pi^2}{4} - \frac{40489}{29400}\biggr) \frac{x^8}{7}
+ \mathcal{O}(x^9)\,,
\label{Mass:Al}
\end{eqnarray}
where after the second equality sign we show the expansion in $x$.
In principle, it is straightforward to obtain exact results in $x$
also the four-loop $C_F T_F^3 n_n n_0^2 \alpha_s^4$ term. However, we
  refrain from presenting such results
because the remaining four-loop color structures are not known.

The remaining three-loop terms can be obtained in a series expansion in $x$
  with the help of the result from~\cite{Bekavac:2007tk}. Including terms up
  to order $x^8$ gives
\begin{eqnarray}
A_F &=& \frac{\pi^2}{3} \biggl(8 a_1 + \frac{13}{4} \pi - \frac{343}{24}\biggr) x
- \biggl(L^2 - \frac{67}{6} L - \frac{17}{8} \pi^2 + \frac{229}{18}\biggr) x^2
+ \frac{\pi^2}{3} \biggl(\frac{11}{3} L + \frac{44}{3} a_1 + \frac{35}{8} \pi - \frac{157}{8}\biggr) x^3
\nonumber\\
&&{} + \biggl[\frac{19}{6} L^3 - \frac{911}{120} L^2
- \biggl(3 \pi^2 a_1 - \frac{3}{2} \zeta_3 - \frac{45}{16} \pi^2 - \frac{40567}{3600}\biggr) L
+ 20 a_4 + \frac{5}{6} a_1^4 + \frac{2}{3} \pi^2 a_1^2 + \frac{11}{16} \pi^2 a_1 + \frac{387}{32} \zeta_3
\nonumber\\
&&\quad{} - \frac{43}{144} \pi^4 - \frac{155}{64} \pi^2 - \frac{2534579}{216000}
\biggr] x^4
+ \frac{7}{5} \pi^2 \biggl(\frac{3}{32} \pi + \frac{1}{5}\biggr) x^5
\nonumber\\
&&{} + \biggl[\frac{1579}{70} L^2 + \biggl(\frac{77}{16} \pi^2 - \frac{328067}{11025}\biggr) L
- \frac{1}{16} \biggl(77 \pi^2 a_1 - \frac{539}{2} \zeta_3 - \frac{83}{15} \pi^2 - \frac{126231437}{1157625}\biggr)
\biggr] \frac{x^6}{9}
- \frac{\pi^2}{28} \biggl(\frac{25}{16} \pi + \frac{1}{7}\biggr) x^7
\nonumber\\
&&{} + \biggl[\frac{2843}{105} L^2 + \biggl(\frac{21}{2} \pi^2 - \frac{718639}{33075}\biggr) \frac{L}{2}
- \frac{1}{4} \biggl(21 \pi^2 a_1 - \frac{147}{2} \zeta_3 - \frac{4379}{240} \pi^2 + \frac{1213332979}{83349000}\biggr)
\biggr] \frac{x^8}{32}
+ \mathcal{O}(x^9)\,,
\nonumber\\
A_A &=& \frac{\pi^2}{8} \biggl(\frac{25}{2} L + \frac{313}{3} a_1 - \frac{13}{3} \pi - \frac{2473}{36}\biggr) x
\nonumber\\
&&{} + \biggl[\frac{L^2}{2} + \biggl(\frac{3}{2} \zeta_3 - \frac{31}{90} \pi^4 + 7 \pi^2 - \frac{7}{3}\biggr) L
- 5 \zeta_5 - \frac{7}{2} \pi^2 \zeta_3 + \frac{79}{4} \zeta_3 - \frac{17}{180} \pi^4 + \frac{35}{18} \pi^2 + \frac{517}{9}
\biggr] \frac{x^2}{4}
\nonumber\\
&&{} + \frac{\pi^2}{24} \biggl(\frac{269}{6} L + \frac{1291}{3} a_1 - \frac{35}{2} \pi - \frac{865}{3}\biggr) x^3
- \biggl[\frac{83}{48} L^3 + \biggl(3 \pi^2 - \frac{3977}{60}\biggr) \frac{L^2}{8}
- \biggl(\frac{3}{2} \pi^2 a_1 - 3 \zeta_3 - \frac{13}{24} \pi^2 - \frac{230293}{28800}\biggr) L
\nonumber\\
&&\quad{} + 10 a_4 + \frac{5}{12} a_1^4 + \frac{\pi^2}{3} a_1^2 + \frac{11}{32} \pi^2 a_1 + \frac{111}{64} \zeta_3
- \frac{161}{1440} \pi^4 - \frac{631}{1152} \pi^2 - \frac{452033}{864000}
\biggr] x^4
+ \frac{\pi^2}{20} \biggl(\frac{79}{9} L - \frac{21}{16} \pi - \frac{2671}{432}\biggr) x^5
\nonumber\\
&&{} + \biggl[\frac{5}{3} L^3 + \frac{9911}{840} L^2 - \biggl(\pi^2 + \frac{8394157}{529200}\biggr) L
+ \frac{1}{12} \biggl(77 \pi^2 a_1 - \frac{509}{2} \zeta_3 - \frac{3607}{60} \pi^2 + \frac{8471770063}{18522000}\biggr)
\biggr] \frac{x^6}{24}
\nonumber\\
&&{} + \frac{\pi^2}{28} \biggl(\frac{57}{25} L + \frac{25}{32} \pi - \frac{11549}{14000}\biggr) x^7
+ \biggl[\frac{43}{27} L^3 + \frac{209}{20} L^2 + \biggl(125 \pi^2 - \frac{12327647}{14700}\biggr) \frac{L}{216}
\nonumber\\
&&\quad{} + \frac{1}{8} \biggl(21 \pi^2 a_1 - \frac{1435}{18} \zeta_3 - \frac{1213519}{45360} \pi^2 + \frac{103012097}{2058000}\biggr)
\biggr] \frac{x^8}{32}
+ \mathcal{O}(x^9)\,,
\nonumber\\
A_h &=& - \biggl(2 L + \frac{13}{5}\biggr) \frac{x^2}{5}
+ \frac{2}{15} \pi^2 x^3
+ \biggl[\frac{3}{70} L^2 + \biggl(\pi^2 - \frac{35887}{4900}\biggr) \frac{L}{3}
- \frac{1}{36} \biggl(13 \pi^2 - \frac{59985349}{514500}\biggr)
\biggr] x^4
\nonumber\\
&&{} - \biggl(244 L^2 - \frac{92779}{315} L + \frac{353877541}{793800}\biggr) \frac{x^6}{945}
- \biggl(47 L^2 + \frac{925823}{13860} L - \frac{4543985839}{384199200}\biggr) \frac{x^8}{770}
+ \mathcal{O}(x^9)\,,
\nonumber\\
A_m &=& - \pi^2 \biggl(\frac{L}{2} - \frac{2}{15}\biggr) x
- \frac{7}{3} x^2
- \frac{5}{6} \pi^2 L x^3
+ \biggl[\frac{2}{3} L^3 - \frac{13}{6} L^2 - \biggl(\frac{\pi^2}{3} - \frac{15}{4}\biggr) L
+ \frac{1}{4} \biggl(\pi^2 + \frac{203}{108}\biggr)
\biggr] x^4
\nonumber\\
&&{} - \biggl(\frac{308}{5} L + \frac{16}{3} \pi^2 - \frac{13159}{225}\biggr) \frac{x^6}{45}
+ \biggl(3 L^2 - \frac{751}{70} L + \frac{2095}{336}\biggr) \frac{x^8}{14}
+ \mathcal{O}(x^9)\,.
\label{Mass:A3}
\end{eqnarray}
\end{widetext}
Starting from three loops the individual terms in Eq.~(\ref{Mass:log}) are
  gauge parameter
dependent.
However, $\xi$ cancels in the three-loop expression for $z'$. It might be that
$z'$ is gauge invariant to all orders, but we have no proof of this
conjecture.

%- }}}
%- {{{ The QED and Bloch--Nordsieck heavy-lepton fields:

\section{The QED and Bloch--Nordsieck heavy-lepton fields}
\label{S:QED}

In QED the matching coefficient $z(\mu)$ is gauge invariant to all orders in $\alpha$~\cite{Grozin:2010wa}.
The proof given in this paper is literally valid only for $n_f = 1$ lepton flavor,
but can be easily generalized for any $n_f$, as we demonstrate in the following.

The QED on-shell renormalization constant $Z_\psi^{\text{os}}$ is gauge
invariant to all orders~\cite{Johnson:1959zz,Zumino:1959wt,Melnikov:2000zc}.
Gauge dependence of the $\overline{\text{MS}}$ $Z_\psi$ can be found using the
so-called LKF transformation~\cite{Landau:1955zz,Fradkin:1955jr}
  for arbitrary $n_f$.  In the gauge where the free photon propagator is
\[
D^0_{\mu\nu}(k) = \frac{1}{k^2} \left(g_{\mu\nu} - \frac{k_\mu k_\nu}{k^2}\right) + \Delta(k) k_\mu k_\nu\,,
\]
the full bare lepton propagator reads
\begin{eqnarray}
&&S(x) = S_L(x) e^{-i e_0^2 (\tilde{\Delta}(x) - \tilde{\Delta}(0))}\,,
\nonumber\\
&&\tilde{\Delta}(x) = \int \frac{d^d k}{(2\pi)^d} \Delta(k) e^{-ikx}\,,
\label{QED:LKF}
\end{eqnarray}
where $S_L(x)$ is the Landau-gauge propagator.
In the covariant gauge $\Delta(k) = (1-\xi_0)/(k^2)^2$,
and $\tilde{\Delta}(0) = 0$ in dimensional regularization.
The lepton fields renormalization does not depend on their masses,
so, let us assume that all $n_f$ flavors are massless.
The propagator has a single Dirac structure
\[
S(x) = S_0(x) e^{\sigma(x)}\,,
\]
where $S_0(x)$ is the $d$-dimensional free propagator.
Then
\[
\sigma(x) = \sigma_L(x) + (1-\xi_0) \frac{e_0^2}{(4\pi)^{d/2}} \left(- \frac{x^2}{4}\right)^\varepsilon \Gamma(-\varepsilon)\,;
\]
re-expressing this result via the renormalized quantities, we obtain
\begin{equation}
\log Z_\psi(\alpha,\xi) = \log Z_L(\alpha) - (1-\xi) \frac{\alpha}{4\pi\varepsilon}\,.
\label{QED: Zpsi}
\end{equation}
In QED $Z_A Z_\alpha = 1$ due to Ward identities, hence
\[
\frac{d\log((1-\xi(\mu)) \alpha(\mu))}{d\log\mu} = - 2 \varepsilon
\]
exactly, and the anomalous dimension
\begin{equation}
\gamma_\psi(\alpha,\xi) = \gamma_L(\alpha) + 2 (1-\xi) \frac{\alpha}{4\pi}
\label{QED:gamma}
\end{equation}
contains $\xi$ only in the one-loop term.

In the Bloch-Nordsieck EFT with $n_l$ light lepton flavors
$Z_h^{\text{os}}$ is gauge-invariant (even if some of these flavors have non-zero masses).
Gauge dependence of the $\overline{\text{MS}}$ $Z_h$ can be found using exponentiation.
The full bare propagator is
\[
S_h(t) = S_{h0}(t) \exp \left(\sum_i w_i\right)\,,
\]
where $w_i$ are webs~\cite{Gatheral:1983cz,Frenkel:1984pz}.
In QED all webs have even numbers of photon legs;
all webs with $>2$ legs are gauge invariant;
all 2-leg webs except the trivial one (the free photon propagator) are gauge invariant, too.
Therefore,
\[
\log \frac{S_h(t)}{S_{hL}(t)} = (1-\xi_0) \frac{e_0^2}{(4\pi)^{d/2}} \left(\frac{it}{2}\right)^{2\varepsilon} \Gamma(-\varepsilon)\,;
\]
re-expressing this result via the renormalized quantities, we obtain
\begin{eqnarray}
&&\log Z_h(\alpha,\xi) = \log Z_{hL}(\alpha) - (1-\xi) \frac{\alpha}{4\pi\varepsilon}\,,
\label{QED: Zh}\\
&&\gamma_h(\alpha,\xi) = \gamma_{hL}(\alpha) + 2 (1-\xi) \frac{\alpha}{4\pi}\,.
\label{QED:gammah}
\end{eqnarray}

Finally, in the abelian case $\zeta_\alpha(\mu) = \zeta_A(\mu)^{-1}$ due to Ward identities,
hence $(1-\xi^{(n_f)}(\mu)) \alpha^{(n_f)}(\mu)) = (1-\xi^{(n_l)}(\mu)) \alpha^{(n_l)}(\mu))$,
and we arrive at the conclusion that $z(\mu)$ is gauge invariant
(some light flavors may be massive, this does not matter).

Let us in the following specify $z(M)$ from Eq~(\ref{QCD:result}) to QED.
Setting $C_F = T_F = d_{FF} = 1$ and $C_A = d_{FA} = 0$
we see that our four-loop result is indeed gauge invariant
and is given by 
\begin{widetext}
\begin{eqnarray}
&&z(M) = 1 - \frac{\alpha}{\pi}
\biggl[1 + \varepsilon \biggl(\frac{\pi^2}{16} + 2\biggr)
    - \varepsilon^2 \biggl(\frac{\zeta_3}{4} - \frac{\pi^2}{12} - 4\biggr)
    - \varepsilon^3 \biggl(\frac{\zeta_3}{3} - \frac{3}{640} \pi^4 - \frac{\pi^2}{6} - 8\biggr)
    + \mathcal{O}(\varepsilon^4) \biggr]
\nonumber\\
&&{} + \left(\frac{\alpha}{\pi}\right)^2
\biggl\{
\pi^2 a_1 - \frac{3}{2} \zeta_3 - \frac{55}{48} \pi^2 + \frac{5957}{1152}
+ \frac{n_l}{12} \biggl(\pi^2 + \frac{113}{8}\biggr)
\nonumber\\
&&\quad{} + \varepsilon \biggl[
- 24 a_4 - a_1^4 - 2 \pi^2 a_1^2 + \frac{31}{4} \pi^2 a_1 - \frac{203}{8} \zeta_3 + \frac{7}{20} \pi^4 - \frac{4903}{1152} \pi^2 + \frac{56845}{6912}
+ n_l \biggl(\zeta_3 + \frac{127}{288} \pi^2 + \frac{851}{192}\biggr)
\biggr]
\nonumber\\
&&\quad{} + \varepsilon^2 \biggl[
- 144 a_5 - 186 a_4 + \frac{6}{5} a_1^5 - \frac{31}{4} a_1^4 + 4 \pi^2 a_1^3 - \frac{31}{2} \pi^2 a_1^2 - \frac{13}{15} \pi^4 a_1 + 30 \pi^2 a_1 + \frac{609}{4} \zeta_5 + \frac{11}{4} \pi^2 \zeta_3 - \frac{28169}{288} \zeta_3
\nonumber\\
&&\qquad{} + \frac{10007}{7680} \pi^4 - \frac{114943}{6912} \pi^2 + \frac{1838165}{41472}
+ \frac{n_l}{24} \biggl(\frac{305}{3} \zeta_3 + \frac{199}{80} \pi^4 + \frac{853}{24} \pi^2 + \frac{5753}{16}\biggr)
\biggr]
+ \mathcal{O}(\varepsilon^3)
\biggr\}
\nonumber\\
&&{} + \left(\frac{\alpha}{\pi}\right)^3
\biggl\{
- 16 a_4 - \frac{2}{3} a_1^4 + \pi^2 a_1^2 + \frac{737}{36} \pi^2 a_1 - \frac{5}{16} \zeta_5 + \frac{\pi^2}{8} \zeta_3 - \frac{4747}{288} \zeta_3 - \frac{13}{360} \pi^4 - \frac{259133}{25920} \pi^2 -\frac{230447}{20736}
\nonumber\\
&&\qquad{} + \frac{n_l}{3} \biggl(16 a_4 + \frac{2}{3} a_1^4 + \frac{4}{3} \pi^2 a_1^2 - \frac{47}{6} \pi^2 a_1 + \frac{137}{8} \zeta_3 - \frac{229}{720} \pi^4 + \frac{139}{24} \pi^2 - \frac{2201}{432}\biggr)
- \frac{n_l^2}{18} \biggl(7 \zeta_3 + \frac{19}{6} \pi^2 + \frac{5767}{432}\biggr)
\nonumber\\
&&\quad{} + \varepsilon \biggl[
- \frac{224}{3} a_5 + 16 \pi^2 a_4 - \frac{5005}{6} a_4 + \frac{28}{45} a_1^5 + \frac{2}{3} \pi^2 a_1^4 - \frac{5005}{144} a_1^4 - \frac{88}{27} \pi^2 a_1^3 - \frac{2}{3} \pi^4 a_1^2 - \frac{11567}{144} \pi^2 a_1^2 + 14 \pi^2 \zeta_3 a_1
\nonumber\\
&&\qquad\quad{} - \frac{2039}{2160} \pi^4 a_1 + \frac{3481}{15} \pi^2 a_1 +\frac{125}{8} \zeta_5 + \frac{29}{32} \zeta_3^2 + \frac{2945}{288} \pi^2 \zeta_3 - \frac{348821}{960} \zeta_3 - \frac{899}{5670} \pi^6 + \frac{64103}{34560} \pi^4 - \frac{224592113}{4147200} \pi^2
\nonumber\\
&&\qquad\quad{} - \frac{2783713}{207360}
+ \frac{n_l}{3} \biggl(224 a_5 + \frac{1124}{3} a_4 - \frac{28}{15} a_1^5 + \frac{281}{18} a_1^4 - \frac{56}{9} \pi^2 a_1^3 + \frac{281}{9} \pi^2 a_1^2 - \frac{17}{90} \pi^4 a_1 - \frac{644}{9} \pi^2 a_1 - \frac{1027}{4} \zeta_5
\nonumber\\
&&\qquad\quad{}
- \frac{119}{16} \pi^2 \zeta_3 + \frac{662}{3} \zeta_3 - \frac{14303}{8640} \pi^4 + \frac{552083}{13824} \pi^2 - \frac{153109}{2592}\biggr)
\nonumber\\
&&\qquad{} - \frac{n_l^2}{54} \biggl(275 \zeta_3 + \frac{23}{5} \pi^4 + \frac{1081}{16} \pi^2 + \frac{253783}{864}\biggr)
\biggr]
+ \mathcal{O}(\varepsilon^2)
\biggr\}
\nonumber\\
&&{} + \left(\frac{\alpha}{\pi}\right)^4
\biggl[
%- 2.162147945818756
L_{\text{QED}}
+\frac{395}{6}a_5
+28\pi^2 a_4
-\frac{58187}{48}a_4
-\frac{79}{144}a_1^5
+\frac{7}{6}\pi^2 a_1^4
-\frac{58187}{1152}a_1^4
-\frac{2411}{216}\pi^2 a_1^3
-\frac{7}{6}\pi^4 a_1^2
\nonumber\\
&&\quad{}
-\frac{69311}{576}\pi^2 a_1^2
+\frac{49}{2}\pi^2 \zeta_3 a_1
-\frac{61}{1728}\pi^4 a_1
+\frac{1414153}{3840}\pi^2 a_1
-\frac{23093}{128}\zeta_5
+\frac{203}{128}\zeta_3^2
+\frac{7771}{576}\pi^2 \zeta_3
-\frac{327897}{640}\zeta_3
-\frac{899}{3240}\pi^6
\nonumber\\
&&\quad{}
+\frac{74911}{69120}\pi^4
-\frac{148407527}{2073600}\pi^2
-\frac{778181617}{9953280}
- n_l (12.18 \pm 0.8)
\nonumber\\
&&\quad{}
- n_l^2 \biggl(\frac{32}{3} a_5 + \frac{188}{9} a_4 - \frac{4}{45} a_1^5 + \frac{47}{54} a_1^4 - \frac{8}{27} \pi^2 a_1^3 + \frac{47}{27} \pi^2 a_1^2 - \frac{31}{270} \pi^4 a_1 - \frac{239}{54} \pi^2 a_1 - \frac{601}{48} \zeta_5 - \frac{\pi^2}{2} \zeta_3 + \frac{6913}{576} \zeta_3
\nonumber\\
&&\qquad{}
- \frac{1297}{51840} \pi^4 + \frac{25729}{10368} \pi^2 - \frac{15877}{165888}\biggr)
+ \frac{n_l^3}{216} \biggl(\frac{467}{2} \zeta_3 + \frac{71}{20} \pi^4 + \frac{167}{3} \pi^2 + \frac{103933}{864}\biggr)
+ \mathcal{O}(\varepsilon)
\biggr]
+ \mathcal{O}(\alpha^5)\,,
\label{QED:result}
\end{eqnarray}
\end{widetext}
where $\alpha = \alpha^{(n_f)}(M)$;
$L_{\text{QED}} = \sum_{i=0,1,2,3,l} L_i$ is the $\varepsilon^0$ term in
$Z_2^{(4)}$ of Eq.~(26)
in~\cite{Laporta:2020fog}. Its numerical value is given in Eq.~(15) in this paper.

Numerically, in pure QED ($n_l=0$) at $\varepsilon=0$ we have
\begin{eqnarray}
&&z(M) = 1 - \frac{\alpha}{\pi}
- 1.09991 \left(\frac{\alpha}{\pi}\right)^2
+ 4.40502 \left(\frac{\alpha}{\pi}\right)^3
\nonumber\\
&&{} - 2.16215 \left(\frac{\alpha}{\pi}\right)^4
+ \mathcal{O}(\alpha^5)\,,
\label{QED:num}
\end{eqnarray}
where $\alpha = \alpha^{(1)}(M)$,
the $\overline{\text{MS}}$ QED coupling with one active flavor
at $\mu=M$, the on-shell electron mass.
In contrast to the QCD case~(\ref{QCD:result})
the coefficients are numerically smaller and have different signs.

%- }}}
%- {{{ Conclusion:

\section{Conclusion}
\label{S:Conc}

We have calculated the (finite) matching coefficient
between the QCD heavy-quark field $Q$
and the corresponding HQET field $h_v$ up to four loops.
Explicit results are presented for $\mu = M$;
results for different values of $\mu$ can be obtained with the help of
(known) renormalization group equations.
The effect of a non-zero light-flavor mass (e.\,g., $c$ in $b$-quark HQET)
is calculated up to three loops.
We also present results for the matching constant in QED.

As a possible application of our results we want
to mention the possibility to obtain the QCD heavy-quark propagator
(say, in Landau gauge)
from lattice QCD results for the HQET propagator.
A heavy-quark field can be put onto the lattice only if $M a \ll 1$,
where $a$ is the lattice spacing.
On the other hand, in HQET simulations there is no lattice $h_v$ field at all.
The HQET propagator is just a straight Wilson line,
i.\,e. a product of lattice gauge links.
It is therefore much easier to obtain the HQET propagator
from lattice simulations.
After taking the continuum limit,
one can get the continuum coordinate-space HQET propagator.
Then the QCD heavy-quark propagator can be obtained
with the help of the matching coefficient $z(\mu)$,
provided that $1/M^n$ corrections can be neglected.
Note that this can be done for arbitrarily heavy QCD quark,
including the case when the use of the dynamic heavy-quark field on the lattice
is impossible.

%- }}}
%- {{{ Acknowledgments:

\section*{Acknowledgments}

We are grateful to R.\,N.~Lee for discussions of the Appendix~\ref{S:2m}.
This research was supported by the Deutsche Forschungsgemeinschaft
(DFG, German Research Foundation) under grant 396021762 --- TRR 257
``Particle Physics Phenomenology after the Higgs Discovery''.
This work was supported in part by the 
EU TMR network SAGEX Marie Sk\l{}odowska-Curie grant agreement No.~764850
and COST action CA16201: Unraveling new physics at the LHC through the precision frontier.
The work of A.\,G.\ was supported
by the Russian Ministry of Science and Higher Education.

%- }}}
%- {{{ Appendix A:

\appendix
\section{The coupling and gluon-field decoupling coefficients}
\label{S:Dec}

The $n_l$-flavor QCD strong coupling constant and gauge parameter
are related to the corresponding quantities in the $n_f$-flavor theory
by the decoupling relations
\begin{eqnarray}
&&\alpha_s^{(n_l)}(\mu) = \zeta_\alpha(\mu) \alpha_s^{(n_f)}(\mu)\,,
\label{Dec:Rel}\\
&&1 - \xi^{(n_l)}(\mu) = \zeta_A(\mu) \left[1 - \xi^{(n_f)}(\mu)\right]\,.
\nonumber
\end{eqnarray}
The decoupling coefficients satisfy the renormalization group equations
\begin{eqnarray}
\frac{d \log \zeta_\alpha(\mu)}{d \log \mu} &=&
2 \bigl[\beta^{(n_f)}(\alpha_s^{(n_f)}(\mu)) - \beta^{(n_l)}(\alpha_s^{(n_l)}(\mu))\bigr]\,,
\nonumber\\
\frac{d \log \zeta_A(\mu)}{d \log \mu} &=&
\gamma_A^{(n_f)}(\alpha_s^{(n_f)}(\mu),\xi^{(n_f)}(\mu))
\nonumber\\
&&{} - \gamma_A^{(n_l)}(\alpha_s^{(n_l)}(\mu),\xi^{(n_l)}(\mu))\,.
\label{Dec:RG}
\end{eqnarray}
It is sufficient to have initial conditions, say, at $\mu = M$ for solving these equations.
For the computation of $z(M)$ we need the decoupling coefficients
up to $\alpha_s^3 \varepsilon$.
Up to the order $\alpha_s^2$ expression exact in $\varepsilon$
can be found in~\cite{Grozin:2012ec}.
The finite three-loop results have been obtained
in~\cite{Chetyrkin:1997un} in term of $N_c$ and
in~\cite{Gerlach:2018hen} for an arbitrary color group.
The $\alpha_s^3 \varepsilon$ terms were derived in the course of four-loop
calculations~\cite{Schroder:2005hy,Chetyrkin:2005ia,Gerlach:2018hen}.
However, results for an arbitrary color group,
including positive powers of $\varepsilon$,
are not explicitly presented in these publications.
Therefore, we present them here:
\begin{widetext}
\begin{eqnarray}
&&\zeta_\alpha(M) = 1
- \frac{\alpha_s}{\pi} T_F n_h \frac{\varepsilon}{9}
\left( \frac{\pi^2}{4} - \zeta_3 \varepsilon + \frac{3}{160} \pi^4 \varepsilon^2 + \mathcal{O}(\varepsilon^3) \right)
\nonumber\\
&&{} - \left(\frac{\alpha_s}{\pi}\right)^2 T_F n_h
\biggl\{ \frac{15}{16} C_F - \frac{2}{9} C_A
+ \frac{\varepsilon}{4} \left[ \frac{C_F}{4} \left( \frac{\pi^2}{3} + \frac{31}{2} \right) + \frac{C_A}{9} \left( \frac{5}{4} \pi^2 + \frac{43}{3} \right) \right]
\nonumber\\
&&\quad{} - \varepsilon^2 \left[ \frac{C_F}{4} \left( \frac{\zeta_3}{3} - \frac{5}{8} \pi^2 - \frac{223}{16} \right) + \frac{C_A}{9} \left( \frac{5}{4} \zeta_3 + \frac{\pi^2}{3} + \frac{523}{72} \right) + \frac{\pi^4}{1296} T_F n_h\right]
+ \mathcal{O}(\varepsilon^3) \biggr\}
\nonumber\\
&&{} + \left(\frac{\alpha_s}{\pi}\right)^3 T_F n_h
\biggl\{ \frac{C_F^2}{3} \left( \pi^2 a_1 - \frac{\zeta_3}{64} - \frac{5}{8} \pi^2 - \frac{77}{192} \right) 
- \frac{C_F C_A}{6} \left( \pi^2 a_1 + \frac{1081}{128} \zeta_3 - \frac{\pi^2}{3} + \frac{8321}{864} \right)
\nonumber\\
&&\qquad{} - \frac{C_A^2}{768} \left( \frac{5}{2} \zeta_3 - \frac{11347}{27} \right)
\nonumber\\
&&\qquad{}
- C_F T_F n_h \left( \frac{7}{64} \zeta_3 + \frac{\pi^2}{9} - \frac{695}{648} \right) 
- \frac{7}{64} C_A T_F n_h \left( \frac{\zeta_3}{2} - \frac{35}{81} \right)
+ \frac{C_F T_F n_l}{18} \left( \pi^2 + \frac{311}{72} \right)
- \frac{C_A T_F n_l}{2592}
\nonumber\\
&&\quad{} - \varepsilon \biggl[ C_F^2 \left( \frac{37}{12} a_4 + \frac{37}{288} a_1^4 + \frac{251}{288} \pi^2 a_1^2 - 2 \pi^2 a_1 + \frac{2759}{576} \zeta_3 - \frac{241}{3456} \pi^4 + \frac{439}{384} \pi^2 + \frac{3329}{3456} \right)
\nonumber\\
&&\qquad{} + C_F C_A \left( \frac{63}{16} a_4 + \frac{21}{128} a_1^4 - \frac{85}{128} \pi^2 a_1^2 + \pi^2 a_1 + \frac{2413}{512} \zeta_3 - \frac{1391}{23040} \pi^4 - \frac{281}{1728} \pi^2 + \frac{451831}{62208} \right)
\nonumber\\
&&\qquad - \frac{C_A^2}{96} \left( 263 a_4 + \frac{263}{24} a_1^4 - \frac{263}{24} \pi^2 a_1^2 + \frac{27347}{288} \zeta_3 - \frac{1687}{1440} \pi^4 - \frac{1063}{216} \pi^2 - \frac{345115}{1944} \right)
\nonumber\\
&&\qquad{}
+ C_F T_F n_h \left( \frac{3}{4} a_4 + \frac{a_1^4}{32} - \frac{\pi^2}{32} a_1^2 - \frac{2}{3} \pi^2 a_1 + \frac{3353}{1152} \zeta_3 - \frac{17}{1920} \pi^4 + \frac{407}{864} \pi^2 - \frac{67037}{15552} \right)
\nonumber\\
&&\qquad{}
+ \frac{C_A T_F n_h}{8} \left( 3 a_4 + \frac{a_1^4}{8} - \frac{\pi^2}{8} a_1^2 + \frac{1799}{864} \zeta_3 - \frac{17}{480} \pi^4 + \frac{113}{1296} \pi^2 + \frac{1165}{11664} \right)
\nonumber\\
&&\qquad{}
- \frac{C_F T_F n_l}{9} \left( \zeta_3 + \frac{403}{192} \pi^2 + \frac{24911}{864} \right)
- \frac{C_A T_F n_l}{27} \left( 5 \zeta_3 + \frac{47}{192} \pi^2 - \frac{6553}{1728} \right) 
\biggr]
+ \mathcal{O}(\varepsilon^2) \biggr\}
+ \mathcal{O}(\alpha_s^4)\,,
\label{Dec:alpha}\\
&&\zeta_A(M) = 1
+ \frac{\alpha_s}{\pi} T_F n_h \frac{\varepsilon}{9}
\left( \frac{\pi^2}{4} - \zeta_3 \varepsilon + \frac{3}{160} \pi^4 \varepsilon^2 + \mathcal{O}(\varepsilon^3) \right)
\nonumber\\
&&{} + \left(\frac{\alpha_s}{\pi}\right)^2 T_F n_h
\biggl\{ \frac{1}{16} \left( 15 C_F - \frac{13}{12} C_A \right)
+ \frac{\varepsilon}{16} \left[ C_F \left( \frac{\pi^2}{3} + \frac{31}{2} \right)
+ \frac{C_A}{12} \left( 5 \pi^2 + \frac{169}{6} \right) \right]
\nonumber\\
&&\qquad{} - \frac{\varepsilon^2}{4}
\biggl[ C_F  \left( \frac{\zeta_3}{3} - \frac{5}{8} \pi^2 - \frac{223}{16} \right) + \frac{C_A}{12} \left( 5 \zeta_3 - \frac{\pi^4}{48} + \frac{13}{24} \pi^2 + \frac{1765}{144} \right)  \biggr]
+ \mathcal{O}(\varepsilon^3) \biggr\}
\nonumber\\
&&{} - \left(\frac{\alpha_s}{\pi}\right)^3 T_F n_h
\biggl\{ \frac{C_F^2}{3} \left( \pi^2 a_1 - \frac{\zeta_3}{64} - \frac{5}{8} \pi^2 - \frac{77}{192} \right)
\nonumber\\
&&\qquad{} - C_F C_A \left( 2 a_4 + \frac{a_1^4}{12} - \frac{\pi^2}{12} a_1^2 + \frac{\pi^2}{6} a_1 + \frac{1765}{768} \zeta_3 - \frac{11}{720} \pi^4 - \frac{\pi^2}{18} + \frac{15977}{20736} \right)
\nonumber\\
&&\qquad{} + C_A^2  \left( a_4 + \frac{a_1^4}{24} - \frac{\pi^2}{24} a_1^2 + \frac{1805}{4608} \zeta_3 - \frac{53}{5760} \pi^4 + \frac{7985}{31104}
- \frac{\xi^{(n_f)}(M)}{48}  \left( \zeta_3 - \frac{677}{144} \right)  \right)
\nonumber\\
&&\qquad{}
- C_F T_F n_h \left( \frac{7}{64} \zeta_3 + \frac{\pi^2}{9} - \frac{695}{648} \right) 
- \frac{C_A T_F n_h}{144}  \left( \frac{287}{8} \zeta_3 - \frac{605}{27} \right)
+ \frac{C_F T_F n_l}{18} \left( \pi^2 + \frac{311}{72} \right)
+ \frac{C_A T_F n_l}{9} \left( \zeta_3 - \frac{665}{432} \right)
\nonumber\\
&&\quad{} - \varepsilon \biggl[ C_F^2  \left( \frac{37}{12} a_4 + \frac{37}{288} a_1^4 + \frac{251}{288} \pi^2 a_1^2 - 2 \pi^2 a_1 + \frac{2759}{576} \zeta_3 - \frac{241}{3456} \pi^4 + \frac{439}{384} \pi^2 + \frac{3329}{3456} \right)
\nonumber\\
&&\qquad{} + C_F C_A \biggl( 12 a_5 + \frac{179}{16} a_4 - \frac{a_1^5}{10} + \frac{179}{384} a_1^4 + \frac{\pi^2}{6} a_1^3 - \frac{371}{384} \pi^2 a_1^2 + \frac{17}{120} \pi^4 a_1 + \pi^2 a_1 - \frac{203}{16} \zeta_5 + \frac{\pi^2}{32} \zeta_3 + \frac{3141}{512} \zeta_3
\nonumber\\
&&\qquad\quad{} - \frac{1057}{7680} \pi^4 - \frac{281}{1728} \pi^2 + \frac{1199393}{124416} \biggr)
\nonumber\\
&&\qquad{} - C_A^2 \biggl( 6 a_5 + \frac{611}{96} a_4 - \frac{a_1^5}{20} + \frac{611}{2304} a_1^4 + \frac{\pi^2}{12} a_1^3 - \frac{611}{2304} \pi^2 a_1^2 + \frac{17}{240} \pi^4 a_1 - \frac{185}{32} \zeta_5 + \frac{3}{128} \pi^2 \zeta_3 + \frac{59395}{27648} \zeta_3
\nonumber\\
&&\qquad\quad{} - \frac{6679}{138240} \pi^4 - \frac{10181}{165888} \pi^2 - \frac{886909}{373248}
+ \frac{\xi^{(n_f)}(M)}{96} \left( 7 \zeta_3 + \frac{\pi^4}{10} - \frac{233}{576} \pi^2 - \frac{5737}{144} \right) \biggr)
\nonumber\\
&&\qquad{}
+ C_F T_F n_h \left( \frac{3}{4} a_4 + \frac{a_1^4}{32} - \frac{\pi^2}{32} a_1^2 - \frac{2}{3} \pi^2 a_1 + \frac{3353}{1152} \zeta_3 - \frac{17}{1920} \pi^4 + \frac{113}{216} \pi^2 - \frac{67037}{15552} \right)
\nonumber\\
&&\qquad{}
+ \frac{C_A T_F n_h}{24} \left( 41 a_4 + \frac{41}{24} a_1^4 - \frac{41}{24} \pi^2 a_1^2 + \frac{5551}{288} \zeta_3 - \frac{697}{1440} \pi^4 - \frac{7}{32} \pi^2 - \frac{4415}{1944} \right)
\nonumber\\
&&\qquad{}
- \frac{C_F T_F n_l}{9} \left( \zeta_3 + \frac{403}{192} \pi^2 + \frac{24911}{864} \right)
- \frac{C_A T_F n_l}{18} \left( \frac{5}{3} \zeta_3 - \frac{\pi^4}{10} + \frac{253}{576} \pi^2 + \frac{27845}{1296} \right)
\biggr] + \mathcal{O}(\varepsilon^2) \biggr\}
+ \mathcal{O}(\alpha_s^4)\,,
\label{Dec:A}
\end{eqnarray}
\end{widetext}
where $\alpha_s = \alpha_s^{(n_f)}(M)$.

%- }}}
%- {{{ Appendix B:

\section{On-shell diagrams with two masses}
\label{S:2m}

Light-quark mass effects in the heavy-quark on-shell propagator diagrams
arise for the first time at two loops, see Fig.~\ref{F:disc}b.
The corresponding integral family can be defined as
\begin{eqnarray}
&&I_{n_1 n_2 n_3 n_4} = C
\raisebox{-5mm}{\begin{picture}(26,12)
\put(13,6){\makebox(0,0){\includegraphics{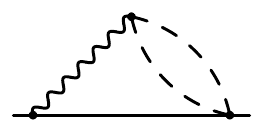}}}
\put(13,-1){\makebox(0,0){1}}
\put(6.5,7.5){\makebox(0,0){2}}
\put(20.5,8.5){\makebox(0,0){3}}
\put(15,3){\makebox(0,0){4}}
\end{picture}}
\nonumber\\
&&{} = \frac{C}{(i \pi^{d/2})^2}
\int \frac{d^d k_1\,d^d k_2}{D_1^{n_1} D_2^{n_2} D_3^{n_3} D_4^{n_4}}\,,\quad
C = \frac{1}{\Gamma^2(1+\varepsilon)}\,,
\nonumber\\
&&D_1 = M^2 - (p + k_1)^2\,,\quad
D_2 = - k_1^2\,,
\nonumber\\
&&D_3 = m^2 - k_2^2\,,\quad
D_4 = m^2 - (k_1 - k_2)^2\,,
\label{2m:Idef}
\end{eqnarray}
with $p^2 = M^2$.
If there are insertions to gluon lines in Fig.~\ref{F:disc}b
containing only massless lines,
such diagrams are expressed via the integrals~(\ref{2m:Idef})
with $n_2 = n + l\varepsilon$,
where $l$ is the total number of loops in these insertions
and $n$ is integer ($n_{1,3,4}$ are always integer).
These integrals have been studied in~\cite{Davydychev:1998si}.
The IBP algorithm obtained there reduces them to four master integrals
\begin{eqnarray}
&&I_{0,l\varepsilon,1,1} = C
\raisebox{-6mm}{\begin{picture}(14,14)
\put(7,7){\makebox(0,0){\includegraphics{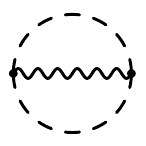}}}
\put(7,9){\makebox(0,0){$l\varepsilon$}}
\end{picture}}\,,\quad
I_{1,l\varepsilon,1,0} = C
\raisebox{-7mm}{\begin{picture}(20,16)
\put(10,7){\makebox(0,0){\includegraphics{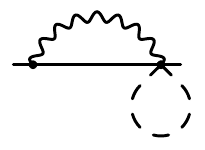}}}
\put(10,15){\makebox(0,0){$l\varepsilon$}}
\end{picture}}\,,
\nonumber\\
&&I_{1,l\varepsilon,1,1} = C
\raisebox{-5mm}{\begin{picture}(26,12)
\put(13,6){\makebox(0,0){\includegraphics{top.pdf}}}
\put(6,7.5){\makebox(0,0){$l\varepsilon$}}
\end{picture}}\,,
\nonumber\\
&&I_{1,1+l\varepsilon,1,1} = C
\raisebox{-5mm}{\begin{picture}(28,14)
\put(17,6){\makebox(0,0){\includegraphics{top.pdf}}}
\put(6.5,7.5){\makebox(0,0){$1+l\varepsilon$}}
\end{picture}}\,.
\label{2m:jdg}
\end{eqnarray}
We set $M=1$ and $m=x$.

It is more convenient to use the column vector
\begin{equation}
j = \bigl(I_{0,l\varepsilon,2,2},I_{2,l\varepsilon,2,0},I_{2,l\varepsilon,2,1},I_{1,l\varepsilon,2,2}\bigr)^T
\label{2m:j}
\end{equation}
as master integrals instead of~(\ref{2m:jdg}).
Differentiating them in $m$ and reducing the results back to $j$~\cite{Kotikov:1990kg},
we obtain the differential equations
\begin{equation}
\frac{d j}{d x} = M(\varepsilon,x) j\,.
\label{2m:DEj}
\end{equation}
In many cases such equations can be reduced
to an $\varepsilon$-form~\cite{Henn:2013pwa}
\begin{equation}
j = T(\varepsilon,x) J\,,\quad
\frac{d J}{d x} = \varepsilon M(x) J\,.
\label{2m:Henn}
\end{equation}
This makes their iterative solution to any order in $\varepsilon$ almost trivial.

Several terms of small-$x$ and large-$x$ expansions of these integrals (with $l=0$)
were obtained in~\cite{Avdeev:1997sz} using the method of regions
(though expressed in a somewhat different language).
Differential equations for on-shell sunsets $I_{n_1,0,n_3,n_4}$ were considered in~\cite{Argeri:2002wz,Grozin:2006xm},
but they were not in $\varepsilon$-form.
Several terms of small-$x$ expansions were obtained from differential equations in~\cite{Onishchenko:2002ri}.
However, the easiest way to obtain any finite number of terms in the small-$x$ and large-$x$ expansions
is neither the method of regions nor differential equations,
but calculating the corresponding residues in the Mellin--Barnes representation~\cite{Davydychev:1998si}.

We use the Mathematica package Libra~\cite{Libra}
which implements the algorithm of~\cite{Lee:2014ioa}
to reduce the master integrals $j$ in Eq.~(\ref{2m:j}) to a canonical basis $J$:
\begin{eqnarray}
j_1 &=& I_{0,l\varepsilon,2,2} = C V_{2,2,l\varepsilon} x^{-2(l+2)\varepsilon}
= \frac{2 (1-(l+1)\varepsilon)}{(l+2) (1-\varepsilon)} J_1\,,
\nonumber\\
j_2 &=& I_{0,l\varepsilon,2,0} = C V_2 M_{2,l\varepsilon} x^{-2\varepsilon}
= \frac{1-2(l+1)\varepsilon}{1-(l+2)\varepsilon} J_2\,,
\nonumber\\
j_3 &=& - \frac{1}{2} \left(J_3 + J_4\right)\,,
\nonumber\\
j_4 &=& I_{1,l\varepsilon,2,2} = \frac{1}{2x}
\biggl\{ - \biggl[1-2x - \frac{2l\varepsilon (1-x)}{1-2\varepsilon}\biggr] J_3
\nonumber\\
&&{} + \biggl[1+2x - \frac{2l\varepsilon (1+x)}{1-2\varepsilon}\biggr] J_4
\biggr\}\,,
\label{2m:T}
\end{eqnarray}
where
\begin{eqnarray}
&&V_{n_1} = \raisebox{-4.5mm}{\begin{picture}(10,11.5)
\put(5,5){\makebox(0,0){\includegraphics{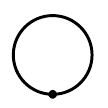}}}
\put(5,10.5){\makebox(0,0){1}}
\end{picture}}
= \frac{\Gamma\bigl(\frac{d}{2}-n_1\bigr)}{\Gamma(n_1)}\,,
\label{2m:V1}\\
&&V_{n_1 n_2 n_3} = \raisebox{-8mm}{\begin{picture}(14,17)
\put(7,8.5){\makebox(0,0){\includegraphics{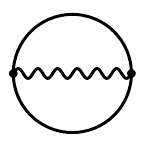}}}
\put(7,1){\makebox(0,0){1}}
\put(7,16){\makebox(0,0){2}}
\put(7,10){\makebox(0,0){3}}
\end{picture}} = H_{0,n_3,n_1,n_2}\,,
%\nonumber\\
%&&\frac{\Gamma\bigl(\frac{d}{2}-n_3\bigr) \Gamma\bigl(n_1+n_3-\frac{d}{2}\bigr) \Gamma\bigl(n_2+n_3-\frac{d}{2}\bigr)}%
%{\Gamma\bigl(\frac{d}{2}\bigr) \Gamma(n_1) \Gamma(n_2)}
%\nonumber\\
%&&\quad{}\times\frac{\Gamma(n_1+n_2+n_3-d)}{\Gamma(n_1+n_2+2n_3-d)}\,,
\label{2m:V2}\\
&&M_{n_1 n_2}
= \raisebox{-4mm}{\begin{picture}(22,9)
\put(11,4.5){\makebox(0,0){\includegraphics{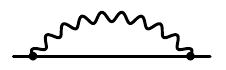}}}
\put(11,1){\makebox(0,0){1}}
\put(11,8){\makebox(0,0){2}}
\end{picture}}
\nonumber\\
&&{} = \frac{\Gamma\bigl(n_1+n_2-\frac{d}{2}\bigr) \Gamma(d-n_1-2n_2)}%
{\Gamma(n_1) \Gamma(d-n_1-n_2)}\,,
\label{2m:M1}
\end{eqnarray}
and~\cite{Grozin:2006xm}
\begin{eqnarray}
&&H_{n_1 n_2 n_3 n_4} =
\raisebox{-5mm}{\begin{picture}(26,12)
\put(13,6){\makebox(0,0){\includegraphics{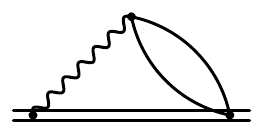}}}
\put(13,-1){\makebox(0,0){1}}
\put(6.5,7.5){\makebox(0,0){2}}
\put(20.5,8.5){\makebox(0,0){3}}
\put(15,3){\makebox(0,0){4}}
\end{picture}} =
\nonumber\\
&&
\frac{\Gamma(n_1/2) \Gamma((n_1-d)/2+n_2+n_3) \Gamma((n_1-d)/2+n_2+n_4)}%
{2 \Gamma(n_1) \Gamma(n_3) \Gamma(n_4)}
\nonumber\\
&&{}\times
\frac{\Gamma(n_1/2+n_2+n_3+n_4-d) \Gamma((d-n_1)/2-n_2)}%
{\Gamma(n_1+2n_2+n_3+n_4-d) \Gamma((d-n_1)/2)}\,.
\label{2m:H2}
\end{eqnarray}

The integrals $J$ satisfy the $\varepsilon$-form differential equations
\begin{equation}
\frac{d J}{d x} = \varepsilon
\left(\frac{M_0}{x} + \frac{M_{+1}}{1-x} + \frac{M_{-1}}{1+x} \right) J\,,
\label{2m:DEJ}
\end{equation}
where
\begin{eqnarray}
&&M_0 =
\left(
\begin{array}{cccc}
- 2 (l+2) &  0  &  0     &  0 \\
  0       & -2  &  0     &  0 \\
  1       & -1  & -(l+2) &  l+2 \\
  1       & -1  &  l+2   & -(l+2)
\end{array}
\right)\,,
\nonumber\\
&&M_{+1} =
\left(
\begin{array}{cccc}
0  &  0 & 0 & 0 \\
0  &  0 & 0 & 0 \\
1  & -1 & 2 & 2 (l+2) \\
0  &  0 & 0 & 0
\end{array}
\right)\,,
\nonumber\\
&&M_{-1} =
\left(
\begin{array}{cccc}
 0 & 0 &  0       &  0 \\
 0 & 0 &  0       &  0 \\
 0 & 0 &  0       &  0 \\
-1 & 1 & -2 (l+2) & -2
\end{array}
\right)\,.
\label{2m:M}
\end{eqnarray}
The first two are, of course, known exactly:
\begin{eqnarray}
J_1 &=& \frac{x^{-2 (l+2) \varepsilon}}{(l+1) \varepsilon^2} \times
\nonumber\\
&&\frac{\Gamma(1 - (l+1) \varepsilon) \Gamma^2(1 + (l+1) \varepsilon) \Gamma(1 + (l+2) \varepsilon)}%
{\Gamma(1-\varepsilon) \Gamma^2(1+\varepsilon) \Gamma(1 + 2 (l+1) \varepsilon)}\,,
\nonumber\\
J_2 &=& \frac{x^{-2\varepsilon}}{(l+1) \varepsilon^2}
\frac{\Gamma(1 - 2 (l+1) \varepsilon) \Gamma(1 + (l+1) \varepsilon)}%
{\Gamma(1 - (l+2) \varepsilon) \Gamma(1+\varepsilon)}\,.
\label{2m:J12}
\end{eqnarray}

The equations for $J_{3,4}$ can be solved iteratively
% ($J_{3,4} = - 2 \sum_{n=0}^\infty L_\mp^{(l,n)} \varepsilon^n$)
in terms of harmonic polylogarithms~\cite{Remiddi:1999ew} of $x$.
However, we need initial conditions.
They can be fixed using the asymptotics of $I_{n_1 n_2 n_3 n_4}$ at $x \to 0$.
It is given by contributions of three regions (Sect.~\ref{S:Mass})
corresponding to residues of the Mellin--Barnes representation~\cite{Davydychev:1998si}
at three series of poles:
\begin{itemize}
\item Hard: the poles $s=-n-n_3-n_4+d/2$ ($n\ge0$ is integer),
the result is a regular series in $x^2$.
The leading term is $C G_{n_3 n_4} M_{n_1,n_2+n_3+n_4-d/2}$, where
\begin{eqnarray}
&&G_{n_1 n_2}
= \raisebox{-6mm}{\begin{picture}(26,13)
\put(13,6.5){\makebox(0,0){\includegraphics{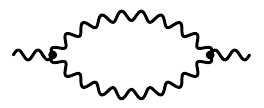}}}
\put(13,12){\makebox(0,0){2}}
\put(13,1){\makebox(0,0){1}}
\end{picture}}
\nonumber\\
&&{} = \frac{\Gamma\bigl(n_1+n_2-\frac{d}{2}\bigr) \Gamma\bigl(\frac{d}{2}-n_1\bigr) \Gamma\bigl(\frac{d}{2}-n_2\bigr)}%
{\Gamma(n_1) \Gamma(n_2) \Gamma(d-n_1-n_2)}\,.
\label{2m:G1}
\end{eqnarray}
\item Sort-hard: $s=-n-n_{3,4}$.
All these poles are double except the first $|n_3-n_4|$ ones
(and hence, the representation of $I_{n_1 n_2 n_3 n_4}$ via hypergeometric functions of $x$ is awkward).
We assume $n_3\ge n_4$, then the result is $x^{d-2n_3}$ times a regular series in $x^2$.
If $n_3>n_4$ then the leading term is $C V_{n_3} M_{n_1,n_2+n_4} x^{d-2n_3}$;
if $n_3=n_4$ there is an extra factor 2 because each of the lines 3, 4 can be soft.
\item Soft: $s=(n_1-d-n)/2+n_2$,
the result is $x^{2(d-n_2-n_3-n_4)-n_1}$ times a regular series in $x$.
The leading term is $H_{n_1 n_2 n_3 n_4} x^{2(d-n_2-n_3-n_4)-n_1}$~(\ref{2m:H2}).
\end{itemize}

For example,
\begin{eqnarray}
&&I_{2,l\varepsilon,2,1} \to \frac{1}{2 \varepsilon^2} \times
\nonumber\\
&&\biggl[ \frac{1}{l+2}
\frac{\Gamma^2(1-\varepsilon) \Gamma(1-2(l+2)\varepsilon) \Gamma(1+(l+2)\varepsilon)}%
{\Gamma(1-2\varepsilon) \Gamma(1-(l+3)\varepsilon) \Gamma(1+\varepsilon)}
\nonumber\\
&&{} - \frac{x^{-2\varepsilon}}{l+1}
\frac{\Gamma(1-2(l+1)\varepsilon) \Gamma(1+(l+1)\varepsilon)}%
{\Gamma(1-(l+2)\varepsilon) \Gamma(1+\varepsilon)}
\nonumber\\
&&{} + \frac{x^{-2(l+2)\varepsilon}}{(l+1) (l+2)} \times
\nonumber\\
&&\frac{\Gamma(1-(l+1)\varepsilon) \Gamma^2(1+(l+1)\varepsilon) \Gamma(1+(l+2)\varepsilon)}%
{\Gamma(1-\varepsilon) \Gamma^2(1+\varepsilon) \Gamma(1+2(l+1)\varepsilon)}
\biggr]\,,
\nonumber\\
\label{2m:asy2021}
\end{eqnarray}
where the 3 contributions are the hard one $C G_{21} M_{2,1+(l+1)\varepsilon}$
(the pole $s=-1-\varepsilon$),
the soft-3 one $C V_2 M_{2,1+l\varepsilon}$ (the pole $s=-1$),
and the soft one $C H_{2,l\varepsilon,2,1} x^{-2(l+2)\varepsilon}$ (the pole $s=-1-(l+1)\varepsilon$).
The leading asymptotics of $I_{1,l\varepsilon,2,2}$ is given by the soft contribution
$C H_{1,l\varepsilon,2,2} x^{-1-2(l+2)\varepsilon}$ (the pole $s=-3/2+(l+1)\varepsilon$):
\begin{eqnarray}
&&I_{1,l\varepsilon,2,2} \to
2^{-1-4(l+2)\varepsilon} \pi^2 x^{-1-2(l+2)\varepsilon}
\frac{1-2(l+1)\varepsilon}{1-2\varepsilon}
\nonumber\\
&&{}\times\frac{\Gamma(1-\varepsilon) \Gamma(1-2(l+1)\varepsilon)}%
{\Gamma(1-2\varepsilon) \Gamma(1-(l+1)\varepsilon)}
\nonumber\\
&&{}\times\frac{\Gamma(1+2(l+1)\varepsilon) \Gamma(1+2(l+2)\varepsilon)}%
{\Gamma^2(1+\varepsilon) \Gamma^2(1+(l+1)\varepsilon) \Gamma(1+(l+2)\varepsilon)}\,.
\label{m0:asy1022}
\end{eqnarray}

Now we can easily obtain any number of expansion terms of $J_{3,4}$ in $\varepsilon$
for $x < 1$ using Libra~\cite{Libra}:
\begin{widetext}
\begin{eqnarray}
J_3 &=& - 2 \biggl\{
H_{1,0}(x) + H_{0,0}(x) + \frac{\pi^2}{3}
+ \biggl[ - (l+2) \bigl(2 H_{1,-1,0}(x) + H_{0,1,0}(x) + H_{0,-1,0}(x)\bigr)
+ 2 H_{1,1,0}(x)
\nonumber\\
&&\quad{} - 2 (l+3) H_{0,0,0}(x)
- l \frac{\pi^2}{6} H_{1}(x) - (l+3) \frac{\pi^2}{3} H_{0}(x)
+ \frac{1}{2} (3l+2) \zeta_3 - (l+2) \pi^2 a_1
\biggr] \varepsilon
\nonumber\\
&&{} + 2 \biggl[
(l+2)^2 \bigl(- 2 H_{1,-1,1,0}(x) + H_{0,1,-1,0}(x) - H_{0,-1,1,0}(x) + H_{0,0,1,0}(x) + H_{0,0,-1,0}(x)\bigr)
\nonumber\\
&&\quad{} + (l+1) (l+2) \bigl(H_{1,0,1,0}(x) + H_{1,0,-1,0}(x)\bigr)
\nonumber\\
&&\quad{} + (l+2) \bigl(- 2 H_{1,1,-1,0}(x) + 2 H_{1,-1,-1,0}(x) - H_{0,1,1,0}(x) + H_{0,-1,-1,0}(x) - 2 H_{1,0,0,0}(x)\bigr)
+ 2 H_{1,1,1,0}(x)
\nonumber\\
&&\quad{} + 2 (l^2+5l+7) H_{0,0,0,0}(x)
+ \frac{\pi^2}{12} \bigl[- 2 l H_{1,1}(x) - (l+2) (5l+6) \bigl(2 H_{1,-1}(x) + H_{0,-1}(x)\bigr)
\nonumber\\
&&\qquad{}
 + (8l^2+23l+12) H_{1,0}(x)
 + l (l+2) H_{0,1}(x) + (6l^2+23l+24) H_{0,0}(x)\bigr]
+ \frac{1}{2} (l+3) (3l+2) \zeta_3 H_{1}(x)
\nonumber\\
&&\quad{} + (l+2) \pi^2 a_1 \bigl[(l+1) H_{1}(x) + (l+2) H_{0}(x)\bigr]
+ (l+2)^2 \pi^2 a_1^2 + (84l^2+227l+122) \frac{\pi^4}{720}
\biggr] \varepsilon^2 \biggr\} + \mathcal{O}(\varepsilon^3)\,,
\nonumber\\
J_4 &=& - 2 \biggl\{
- H_{-1,0}(x) + H_{0,0}(x) - \frac{\pi^2}{6}
+ \biggl[ - (l+2) \bigl(2 H_{-1,1,0}(x) - H_{0,-1,0}(x) - H_{0,1,0}(x)\bigr)
+ 2 H_{-1,-1,0}(x)
\nonumber\\
&&\quad{} - 2 (l+3) H_{0,0,0}(x)
- (5l+6) \frac{\pi^2}{6} H_{-1}(x) + (2l+3) \frac{\pi^2}{3} H_{0}(x)
+ \frac{1}{2} (3l+2) \zeta_3 + (l+2) \pi^2 a_1
\biggr] \varepsilon
\nonumber\\
&&{} + 2 \biggl[
(l+2)^2 \bigl(2 H_{-1,1,-1,0}(x) + H_{0,-1,1,0}(x) - H_{0,1,-1,0}(x) - H_{0,0,-1,0}(x) - H_{0,0,1,0}(x)\bigr)
\nonumber\\
&&\quad{} + (l+1) (l+2) \bigl(H_{-1,0,-1,0}(x) + H_{-1,0,1,0}(x)\bigr)
\nonumber\\
&&\quad{} + (l+2) \bigl(2 H_{-1,-1,1,0}(x) - 2 H_{-1,1,1,0}(x) - H_{0,-1,-1,0}(x) + H_{0,1,1,0}(x) + 2 H_{-1,0,0,0}(x)\bigr)
- 2 H_{-1,-1,-1,0}(x)
\nonumber\\
&&\quad{} + 2 (l^2+5l+7) H_{0,0,0,0}(x)
+ \frac{\pi^2}{12} \bigl[2 (5l+6) H_{-1,-1}(x) + l (l+2) \bigl(2 H_{-1,1}(x) - H_{0,1}(x)\bigr)
\nonumber\\
&&\qquad{} 
+ (4l^2+13l+12) H_{-1,0}(x)
+ (l+2) (5l+6) H_{0,-1}(x) - (2l+3) (3l+8) H_{0,0}(x)\bigr]
- \frac{1}{2} (l+3) (3l+2) \zeta_3 H_{-1}(x)
\nonumber\\
&&\quad{} + (l+2) \pi^2 a_1 \bigl[(l+1) H_{-1}(x) - (l+2) H_{0}(x)\bigr]
- (l+2)^2 \pi^2 a_1^2 - (36l^2+103l+58) \frac{\pi^4}{720}
\biggr] \varepsilon^2 \biggr\} + \mathcal{O}(\varepsilon^3)\,.
\label{2m:J34}
\end{eqnarray}
\end{widetext} 
Up to order $\varepsilon^1$ all harmonic polylogarithms
can be transformed to logarithms and ordinary
polylogarithms up to $\Li3$, e.\,g., using the Mathematica package
HPL~\cite{Maitre:2005uu,Maitre:2007kp}.

Next we consider the case $x>1$.
We can re-write the differential equation~(\ref{2m:DEJ}) in the form
\begin{eqnarray}
&&\frac{d J}{d x^{-1}} = \varepsilon \times{}
\label{2m:DE2}\\
&&\left( \frac{- M_0 + M_{+1} - M_{-1}}{x^{-1}} + \frac{M_{+1}}{1 - x^{-1}} + \frac{M_{-1}}{1 + x^{-1}}
\right) J\,.
\nonumber
\end{eqnarray}
It can be solved in terms of harmonic polylogarithms of $x^{-1}$,
this is convenient for $x>1$.
We use the asymptotics $x \to +\infty$ for boundary conditions.
There are 2 regions:
\begin{itemize}
\item All lines in~(\ref{2m:Idef}) are hard (momenta of order $m$).
This corresponds to the series of right poles in the Mellin--Barnes representation
$s=n+n_1+n_2-d/2$,
i.\,e., to the first term in the hypergeometric representation~(A1) in~\cite{Davydychev:1998si},
and gives $x^{2(d-n_1-n_2-n_3-n_4)}$ times a regular series in $x^{-2}$.
The leading contribution to $I_{n_1 n_2 n_3 n_4}$
is $C V_{n_3,n_4,n_1+n_2} x^{2(d-n_1-n_2-n_3-n_4)}$.
\item Lines 1, 2 are soft (momenta of order $M$).
This corresponds to right poles at $s=n$,
i.\,e., to the second hypergeometric term,
and gives $x^{d-2(n_3+n_4)}$ times a regular series in $x^{-2}$.
The leading asymptotics is $C M_{n_1 n_2} V_{n_3+n_4} x^{d-2(n_3+n_4)}$.
\end{itemize}
For $I_{2,l\varepsilon,2,1}$ these two contributions are $\sim x^{-2-2\varepsilon}$ and $\sim x^{-2-(l+2)\varepsilon}$;
for $I_{1,l\varepsilon,2,2}$ the leading contribution is hard, $\sim x^{-2-(l+1)\varepsilon}$.
This information is sufficient for solving the differential equations for $x>1$ using Libra~\cite{Libra}:
\begin{widetext}
\begin{eqnarray}
J_3 &=& - 2 \biggl\{ - H_{1,0}(x^{-1})
+ \biggl[- (l+4) H_{0,1,0}(x^{-1}) - 2 (l+3) H_{1,0,0}(x^{-1})
+ (l+2) \bigl(2 H_{1,-1,0}(x^{-1}) + H_{0,-1,0}(x^{-1})\bigr)
\nonumber\\
&&\quad{} - 2 H_{1,1,0}(x^{-1})
+ l \frac{\pi^2}{6} H_{1}(x^{-1})\biggr] \varepsilon
\nonumber\\
&&{} + 2 \biggl[(l+2)^2 \bigl(2 H_{1,-1,1,0}(x^{-1}) + H_{0,-1,1,0}(x^{-1})\bigr)
+ (l+2) (l+5) H_{1,0,-1,0}(x^{-1})
\nonumber\\
&&\quad{} + (l+2) (l+4) \bigl(H_{0,1,-1,0}(x^{-1}) + H_{0,0,-1,0}(x^{-1})\bigr)
+ (l+2) (l+3) \bigl(2 H_{1,-1,0,0}(x^{-1}) + H_{0,-1,0,0}(x^{-1})\bigr)
\nonumber\\
&&\quad{} - (l+3) (l+4) H_{0,1,0,0}(x^{-1})
- (l^2+6l+10) H_{0,0,1,0}(x^{-1}) - (l^2+5l+8) H_{1,0,1,0}(x^{-1})
\nonumber\\
&&\quad{} - 2 (l^2+5l+7) H_{1,0,0,0}(x^{-1})
- (l+4) H_{0,1,1,0}(x^{-1}) - 2 (l+3) H_{1,1,0,0}(x^{-1})
\nonumber\\
&&\quad{} + (l+2) \bigl(2 H_{1,1,-1,0}(x^{-1}) - 2 H_{1,-1,-1,0}(x^{-1}) - H_{0,-1,-1,0}(x^{-1})\bigr)
- 2 H_{1,1,1,0}(x^{-1})
\nonumber\\
&&\quad{} + l \frac{\pi^2}{12} \bigl[(l+4) H_{0,1}(x^{-1})
- (l+2) \bigl(2 H_{1,-1}(x^{-1}) + H_{0,-1}(x^{-1})\bigr)
+ 2 H_{1,1}(x^{-1}) + H_{1,0}(x^{-1})\bigr]
\biggr] \varepsilon^2 \biggr\} + \mathcal{O}(\varepsilon^3)\,,
\nonumber\\
J_4 &=& - 2 \biggl\{ H_{-1,0}(x^{-1})
+ \biggl[(l+4) H_{0,-1,0}(x^{-1}) + 2 (l+3) H_{-1,0,0}(x^{-1})
+ (l+2) \bigl(2 H_{-1,1,0}(x^{-1}) - H_{0,1,0}(x^{-1})\bigr)
\nonumber\\
&&\quad{} - 2 H_{-1,-1,0}(x^{-1})
- l \frac{\pi^2}{6} H_{-1}(x^{-1})\biggr] \varepsilon
\nonumber\\
&&{} + 2 \biggl[(l+2)^2 \bigl(- 2 H_{-1,1,-1,0}(x^{-1}) + H_{0,1,-1,0}(x^{-1})\bigr)
+ (l+2) (l+5) H_{-1,0,1,0}(x^{-1})
\nonumber\\
&&\quad{} + (l+2) (l+4) \bigl(H_{0,-1,1,0}(x^{-1}) - H_{0,0,1,0}(x^{-1})\bigr)
+ (l+2) (l+3) \bigl(2 H_{-1,1,0,0}(x^{-1}) - H_{0,1,0,0}(x^{-1})\bigr)
\nonumber\\
&&\quad{} + (l+3) (l+4) H_{0,-1,0,0}(x^{-1})
+ (l^2+6l+10) H_{0,0,-1,0}(x^{-1}) - (l^2+5l+8) H_{-1,0,-1,0}(x^{-1})
\nonumber\\
&&\quad{} + 2 (l^2+5l+7) H_{-1,0,0,0}(x^{-1})
- (l+4) H_{0,-1,-1,0}(x^{-1}) - 2 (l+3) H_{-1,-1,0,0}(x^{-1})
\nonumber\\
&&\quad{} - (l+2) \bigl(2 H_{-1,-1,1,0}(x^{-1}) - 2 H_{-1,1,1,0}(x^{-1}) + H_{0,1,1,0}(x^{-1})\bigr)
+ 2 H_{-1,-1,-1,0}(x^{-1})
\nonumber\\
&&\quad{} + l \frac{\pi^2}{12} \bigl[- (l+4) H_{0,-1}(x^{-1})
- (l+2) \bigl(2 H_{-1,1}(x^{-1}) - H_{0,1}(x^{-1})\bigr)
+ 2 H_{-1,-1}(x^{-1}) - H_{-1,0}(x^{-1})\bigr]
\biggr] \varepsilon^2 \biggr\}
\nonumber\\
&&{} + \mathcal{O}(\varepsilon^3)\,.
\label{2m:J34a}
\end{eqnarray}
\end{widetext}
This is, of course, the analytical continuation of~(\ref{2m:J34}) to $x>1$.
The same results~(\ref{2m:J34a}) can be obtained if we express $J_{3,4}$ via $I_{2,l\varepsilon,2,1}$ and $I_{1,l\varepsilon,2,2}$ using~(\ref{2m:T})
and expand the hypergeometric representations (see Eq.~(A1) in~\cite{Davydychev:1998si}) of these two integrals
in $\varepsilon$ using the Mathematica package HypExp~\cite{Huber:2005yg,Huber:2007dx}.
However, solving the differential equations~(\ref{2m:DE2}) up to higher orders in $\varepsilon$
is simpler than expanding hypergeometric functions.

Both~(\ref{2m:J34}) and~(\ref{2m:J34a}) lead to identical results at $x=1$:
\begin{eqnarray}
&&J_3(1) = - \frac{\pi^2}{3} + \frac{1}{2} (l+2) \bigl(2 \pi^2 a_1 - 7 \zeta_3\bigr) \varepsilon
\nonumber\\
&&{} - \biggl[ (l+2) (l+3) \biggl(8 a_4 + \frac{1}{3} a_1^4 + \frac{2}{3} \pi^2 a_1^2\biggr)
\nonumber\\
&&\quad{} + (17l^2-36l-124) \frac{\pi^4}{360}
\biggr] \varepsilon^2 + \mathcal{O}(\varepsilon^3)\,,
\nonumber\\
&&J_4(1) = \frac{\pi^2}{6} - \frac{1}{2} \bigl(2 \pi^2 a_1 - 7 \zeta_3\bigr) \varepsilon
\nonumber\\
&&{} + \biggl[ (l+3) \biggl(8 a_4 + \frac{1}{3} a_1^4 + \frac{2}{3} \pi^2 a_1^2\biggr)
\nonumber\\
&&\quad{} + (24l^2+27l-62) \frac{\pi^4}{360}
\biggr] \varepsilon^2 + \mathcal{O}(\varepsilon^3)\,.
\label{2m:x1}
\end{eqnarray}
If $l=0$ and $x=1$, we obviously have $I_{1022}(1) = I_{2021}(1)$, and hence
\begin{equation}
J_3(1) = - 2 J_4(1) = - 4 I_{2021}(1)\,.
\label{2m:l0x1}
\end{equation}
Expanding the hypergeometric representation~\cite{Davydychev:1998si}
of $I_{2021}$ (or $I_{1022}$) at $x=1$ in $\varepsilon$ we get~(\ref{2m:x1}) with $l=0$.
Alternatively, we can use another hypergeometric representation~\cite{Broadhurst:1991fi,Broadhurst:1996az}.
Using integration by parts we obtain
\begin{eqnarray}
&&I_{2021}(1) = \frac{7}{32 \varepsilon^2} \biggl[
\frac{\Gamma(1-\varepsilon) \Gamma^2(1+2\varepsilon) \Gamma(1+3\varepsilon)}%
{\Gamma^2(1+\varepsilon) \Gamma(1+4\varepsilon)}
- 1\biggr]
\nonumber\\
&&{} + \frac{2^{-2-6\varepsilon} \pi^2}{3}
\frac{\Gamma^3(1+2\varepsilon) \Gamma(1+3\varepsilon)}%
{\Gamma^5(1+\varepsilon) \Gamma^2(1+2\varepsilon)}
+ \frac{3}{4} \varepsilon^2 B_4(\varepsilon)\,,
\label{2m:B4}
\end{eqnarray}
where $B_4(\varepsilon)$ is given by the formulas~(41), (43) in~\cite{Broadhurst:1996az}.
This leads to the same result.

The functions $L_\mp(x) = - \frac{1}{2} J_{3,4}(l=0,\varepsilon=0)$
were used in~\cite{Broadhurst:1991fy,Broadhurst:1991fi,Broadhurst:1994se,Davydychev:1998si}.
In addition to the two expressions for these functions in~(\ref{2m:J34}) and~(\ref{2m:J34a}),
several additional representations can be found in~\cite{Davydychev:1998si}.

The results~(\ref{2m:J34}) and~(\ref{2m:J34a}) are expansions in $\varepsilon$
where the coefficients are exact functions of $x$.
On the other hand, it is straightforward to obtain expansions of $J_{3,4}$ in $x$ (or $x^{-1}$)
to any finite order using residues of left (or right) poles in the Mellin--Barnes representations
of the integrals $j_{3,4}$~(\ref{2m:j}),
the coefficients being exact functions of $\varepsilon$.
If we expand them in $\varepsilon$,
they should agree with expansions of~(\ref{2m:J34}) in $x$ and of~(\ref{2m:J34a}) in $x^{-1}$.
We have checked this up to rather high degrees of $x$ and $x^{-1}$.

Now we can find all contributions to $Z_j^{\text{os}}$ ($j=M$, $Q$)
with the maximum number of quark loops,
at most one of which is massive, to all orders exactly in $\varepsilon$:
\begin{eqnarray}
&&Z_j^{\text{os}} = 1 + C_F \sum_l T_F^{l-1} (n_0 P)^{l-2}
\Bigl[n_0 P B_{j0}^{(l)}
\label{2m:Z}\\
&&{} + (l-1) \sum_i B_j^{(l)}(x_i)\Bigr]
\left(\frac{g_0^2 M^{-2\varepsilon}}{(4\pi)^{d/2}} \Gamma(\varepsilon)\right)^l + \cdots\,,
\nonumber
\end{eqnarray}
where $g_0 \equiv g_0^{(n_f)}$,
$n_0$ is the number of massless flavors,
the sum runs over all massive flavors with $x_i = m_i/M$
(including the external flavor with $x=1$)
and dots refer to other color structures.
Here
\begin{widetext}
\begin{eqnarray}
&&P = - 4 \frac{1-\varepsilon}{(1-2\varepsilon) (3-2\varepsilon)}
\frac{\Gamma^2(1-\varepsilon)}{\Gamma(1-2\varepsilon)}\,,
\nonumber\\
&&B_{M0}^{(l)} = - 2 \frac{(3-2\varepsilon) (1-l\varepsilon)}{l (1-(l+1)\varepsilon) (2-(l+1)\varepsilon)}
\frac{\Gamma(1+l\varepsilon) \Gamma(1-2l\varepsilon)}{\Gamma(1+\varepsilon) \Gamma(1-(l+1)\varepsilon)}\,,\quad
B_{Q0}^{(l)} = B_{M0}^{(l)} (1+(l-1)\varepsilon)\,,
\nonumber\\
&&B_M^{(l)}(x) = 2 p_0 \biggl\{
- 2 \frac{1-\varepsilon}{l} \biggl[1-l\varepsilon + l\varepsilon \frac{1-(l-1)\varepsilon}{1+(l-1)\varepsilon} x^2\biggr] J_1^{(l-2)}(x)
+ \biggl[p_1 + 2 \varepsilon \frac{1 + (2l-3)\varepsilon - (l-1) (l+2) \varepsilon^2}{1+(l-1)\varepsilon} x^2\biggr] J_2^{(l-2)}(x)
\nonumber\\
&&{} - (p_1 (1+x^2) + p_2 x) (1-x)^2 J_3^{(l-2)}(x)
- (p_1 (1+x^2) - p_2 x) (1+x)^2 J_4^{(l-2)}(x)
\biggr\}\,,
\nonumber\\
&&B_Q^{(l)}(x) = p_0 \biggl\{
- \frac{2 \varepsilon}{l (1-\varepsilon) (1+2(l-1)\varepsilon) (3+2(l-1)\varepsilon)}
\nonumber\\
&&{}\times\biggl[(1+(l-1)\varepsilon)
(19l-3-(11l^2+50l-11)\varepsilon-2(4l^3-20l^2-15l+6)\varepsilon^2+4(4l^3-11l^2+2l+1)\varepsilon^3-8l(l-1)^2\varepsilon^4)
\nonumber\\
&&\quad{} + \frac{l (1-\varepsilon) (1+2(l-1)\varepsilon) (3+2(l-1)\varepsilon)}{1+(l-1)\varepsilon}
(4-(3l+1)\varepsilon-(l-1)(l-7)\varepsilon^2-4(l-1)\varepsilon^3) x^2
\biggr] J_1^{(l-2)}(x)
\nonumber\\
&&{} + 2 \biggl[2 (1-\varepsilon) (1+(l-1)\varepsilon) (1-l\varepsilon)
+ \varepsilon \frac{4+(11l-15)\varepsilon-(l-1)(l+17)\varepsilon^2-2(l-1)(l^2-3)\varepsilon^3}{1+(l-1)\varepsilon} x^2
\biggr] J_2^{(l-2)}(x)
\nonumber\\
&&{} + (p_3 + p_4 x + p_5 x^2 + p_6 x^3) (1-x) J_3^{(l-2)}(x)
+ (p_3 - p_4 x + p_5 x^2 - p_6 x^3) (1+x) J_4^{(l-2)}(x) \biggr\}\,,
\label{2m:B}
\end{eqnarray}
\end{widetext}
where
\begin{eqnarray*}
&&p_0 = \frac{2 \varepsilon^2}{(1-2\varepsilon) (1-(l+1)\varepsilon) (2-(l+1)\varepsilon)}\,,\\
&&p_1 = 2 (1-\varepsilon) (1-l\varepsilon)\,,\\
&&p_2 = \frac{(1-(l+1)\varepsilon) (2+(l-3)\varepsilon-2(l-1)\varepsilon^2)}{1+(l-1)\varepsilon}\,,\\
&&p_3 = - 4 (1-\varepsilon) (1+(l-1)\varepsilon) (1-l\varepsilon)\,,\\
&&p_4 = \frac{2+(3l-5)\varepsilon-(l-1)(5l-1)\varepsilon^2+4(l-1)^2\varepsilon^3(2-\varepsilon)}{1+(l-1)\varepsilon}\,,\\
&&p_5 = [2-(5l+13)\varepsilon+(l^2-6l+29)\varepsilon^2\\
&&\quad{}+2(l^3-l^2+9l-13)\varepsilon^3-8(l-1)\varepsilon^4]/[1+(l-1)\varepsilon]\,,\\
&&p_6 = 4 (1-\varepsilon) (3-2\varepsilon) (1-l\varepsilon)\,.
\end{eqnarray*}
The results~(\ref{2m:B}) at $l=2$ agree with~\cite{Davydychev:1998si} exactly in $\varepsilon$.
Note that
\begin{equation}
\lim_{x\to0} B_M^{(l)}(x) = B_{M0}^{(l)} P\,,
\label{2m:BMlim}
\end{equation}
so that $Z_M^{\text{os}}$ has a smooth limit $x\to0$;
this is not so for $Z_Q^{\text{os}}$.

\begin{figure}[ht]
\begin{picture}(60,32)
\put(25,16){\makebox(0,0){\includegraphics[width=5cm]{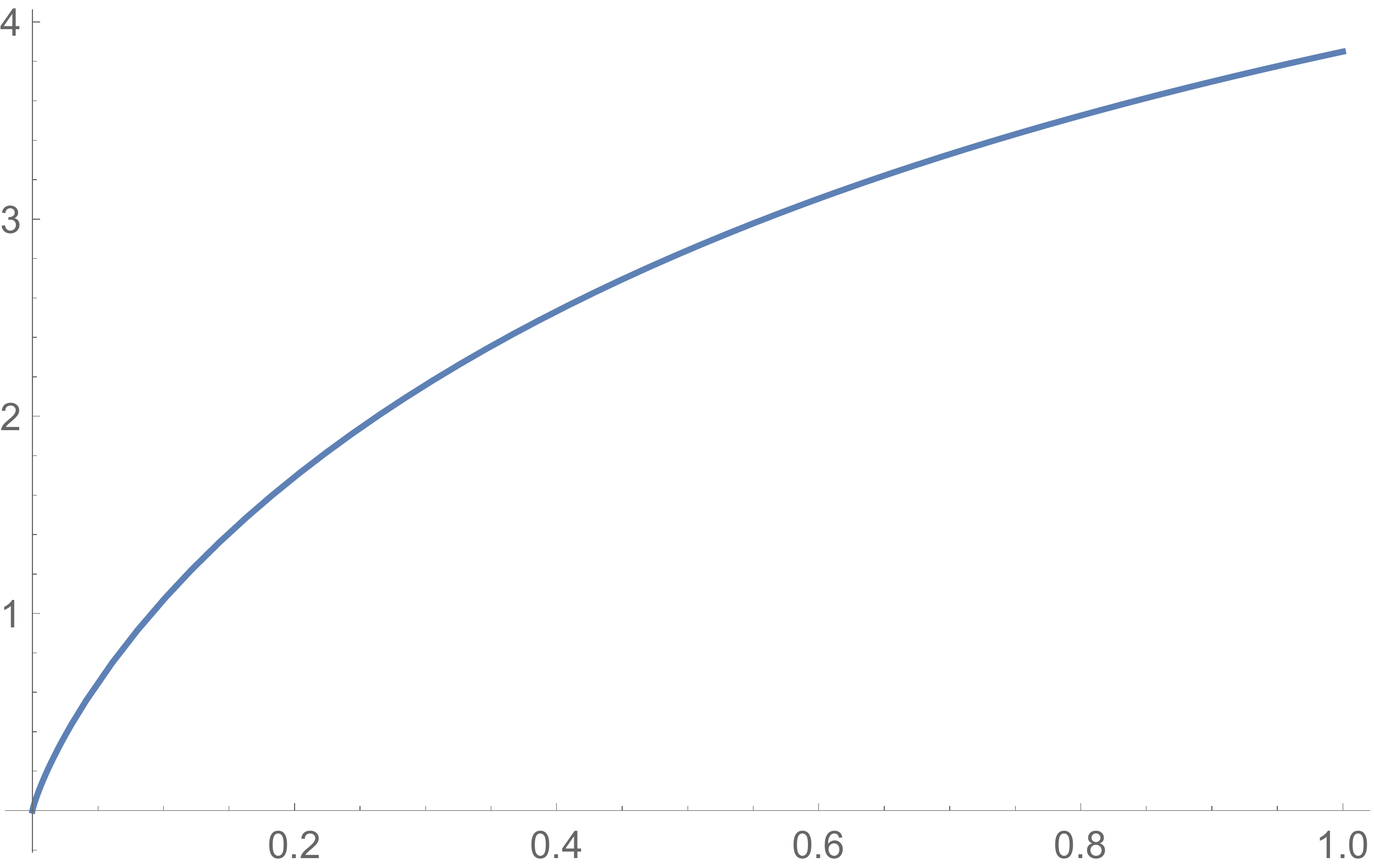}}}
\put(55,1.2){\makebox(0,0){$x$}}
\end{picture}
\caption{The coefficient of $(\alpha_s/\pi)^3 C_F T_F^2 n_0$ in $\Delta_m(x)$.}
\label{F:zm}
\end{figure}

The contribution of these color structures to the ratio of the $\overline{\text{MS}}$ mass
and the on-shell one $z_m(\mu) = M(\mu)/M$ can be written as
\begin{equation}
z_m(M) = z_m^{(\beta_0)} + \sum_i \Delta_m(x_i) + \cdots\,,
\label{2m:zm}
\end{equation}
where $z_m^{(\beta_0)}$ is the well-known large-$\beta_0$ result~\cite{Beneke:1994sw}
\begin{eqnarray}
&&z_m^{(\beta_0)} = 1 + \frac{1}{2} \int_0^b \frac{d b}{b}
\left(\frac{\gamma_m(b)}{b} - \frac{\gamma_{m0}}{\beta_0}\right)
\nonumber\\
&&\quad{} + \frac{1}{\beta_0} \int_0^\infty d u\,S(u)\,e^{-u/b}\,,
\nonumber\\
&&\gamma_m(b) = \frac{2}{3} C_F \frac{b}{\beta_0}
\frac{(3+2b) \Gamma(4+2b)}{\Gamma(3+b) \Gamma^2(2+b) \Gamma(1-b)}\,,
\nonumber\\
&&\gamma_{m0} = 6 C_F\,,
\nonumber\\
&&S(u) = - 6 C_F \left[e^{(5/3)u} \frac{\Gamma(u) \Gamma(1-2u)}{\Gamma(3-u)} (1-u) - \frac{1}{2u}\right]\,,
\nonumber\\
&&b = \beta_0 \frac{\alpha_s}{4\pi}\quad
(\alpha_s \equiv \alpha_s^{(n_f)}(M))
\,.
\label{2m:zmbeta0}
\end{eqnarray}
Note that we first expand $S(u)$ in $u$, then integrate term-by-term assuming $\beta_0>0$,
and at the very end substitute $\beta_0 \to -(4/3) T_F n_f$.
$\Delta_m(x)$
comes from the differences of diagrams with a single massive quark loop
and corresponding diagrams with all quark loops being massless and is
  given by
\begin{widetext}
\begin{eqnarray}
&&\Delta_m(x) = C_F T_F \left(\frac{\alpha_s}{\pi}\right)^2 \biggl\{ \frac{1}{2}
\biggl[(1-x)^2 (1+x+x^2) H_{1,0}(x) - (1+x)^2 (1-x+x^2) H_{-1,0}(x)
+ 2 x^4 H_{0,0}(x)
\nonumber\\
&&\quad{} + x^2 H_{0}(x) - x (3+3x^2-x^3) \frac{\pi^2}{6} + \frac{3}{2} x^2\biggr]
\nonumber\\
&&{} + T_F n_0 \frac{\alpha_s}{\pi} \frac{2}{3}
\biggl[(1-x)^2 (1+x+x^2) \left(H_{1,-1,0}(x) + \frac{\pi^2}{12} H_{1}(x)\right)
\nonumber\\
&&\quad{} + (1+x)^2 (1-x+x^2) \left(H_{-1,1,0}(x) + \frac{5 \pi^2}{12} H_{-1}(x)\right)
- x (1+x^2) \left(H_{0,1,0}(x) + H_{0,-1,0}(x) + \pi^2 a_1\right)
\nonumber\\
&&\quad{} + x^4 \left(2 H_{0,0,0}(x) - \frac{13}{6} H_{0,0}(x) - \frac{3}{2} \zeta_3\right)
- x (3+3x^2+x^3) \frac{\pi^2}{6} H_{0}(x)
- \frac{1}{12} (1-x)^2 (13+10x+13x^2) H_{1,0}(x)
\nonumber\\
&&\quad{} + \frac{1}{12} (1+x)^2 (13-10x+13x^2) H_{-1,0}(x)
+ x (48-12x+48x^2-13x^3) \frac{\pi^2}{72}
- \frac{7}{12} x^2 H_{0}(x)
- \frac{11}{8} x^2
\biggr]
\nonumber\\
&&{} + \left(T_F n_0 \frac{\alpha_s}{\pi}\right)^2 \frac{2}{3}
\biggl[ (1-x)^2 (1+x+x^2)
\nonumber\\
&&\qquad{}\times\biggl(- 2 H_{1,-1,1,0}(x) + H_{1,0,1,0}(x) + H_{1,0,-1,0}(x)
- \frac{\pi^2}{6} \left(5 H_{1,-1}(x) - 4 H_{1,0}(x)\right)
+ \left(\pi^2 a_1 + \frac{3}{2} \zeta_3\right) H_{1}(x)\biggr)
\nonumber\\
&&\quad{} + (1+x)^2 (1-x+x^2)
\nonumber\\
&&\qquad{}\times\left(2 H_{-1,1,-1,0}(x) + H_{-1,0,-1,0}(x) + H_{-1,0,1,0}(x)
+ \frac{\pi^2}{6} \left(H_{-1,1}(x) + 2 H_{-1,0}(x)\right)
+ \left(\pi^2 a_1 - \frac{3}{2} \zeta_3\right) H_{-1}(x)\right)
\nonumber\\
&&\quad{} + 2 x (1+x^2) \biggl(- H_{0,1,-1,0}(x) + H_{0,-1,1,0}(x) - H_{0,0,1,0}(x) - H_{0,0,-1,0}(x)
+ \frac{4}{3} \left(H_{0,1,0}(x) + H_{0,-1,0}(x) + \pi^2 a_1\right)
\nonumber\\
&&\qquad{} - \frac{\pi^2}{12} \left(H_{0,1}(x) - 5 H_{0,-1}(x) + 6 H_{0,0}(x)\right)
- \pi^2 a_1 H_{0}(x) - \pi^2 a_1^2\biggr)
\nonumber\\
&&\quad{} + x^4 \left(4 H_{0,0,0,0}(x) - \frac{13}{3} H_{0,0,0}(x) + \frac{89}{36} H_{0,0}(x)\right)
\nonumber\\
&&\quad{} - \frac{1}{6} (1-x)^2 (13+10x+13x^2) \left(\!H_{1,-1,0}(x) + \frac{\pi^2}{12} H_{1}(x)\right)
- \frac{1}{6} (1+x)^2 (13-10x+13x^2) \left(\!H_{-1,1,0}(x) + \frac{5 \pi^2}{12} H_{-1}(x)\right)
\nonumber\\
&&\quad{} + \frac{1}{72} (1-x)^2 (89+68x+89x^2) H_{1,0}(x)
- \frac{1}{72} (1+x)^2 (89-68x+89x^2) H_{-1,0}(x)
\nonumber\\
&&\quad{} + x (48+6x+48x^2+13x^3) \frac{\pi^2}{36} H_{0}(x)
+ \frac{47}{72} x^2 H_{0}(x)
+ \frac{1}{4} x^2 (6+13x^2) \zeta_3
- x (5+5x^2-2x^3) \frac{\pi^4}{30}
\nonumber\\
&&\quad{} - x (330-192x+330x^2-89x^3) \frac{\pi^2}{432}
+ \frac{33}{16} x^2 \biggr]
+ \mathcal{O}(\alpha_s^3) \biggr\}\,.
\label{2m:Delta}
\end{eqnarray}
\end{widetext}
Note that $\Delta_m(0) = 0$.
Expanding the three-loop term in $x$ we reproduce the series (up to $x^8$) obtained in~\cite{Bekavac:2007tk}.
The three-loop coefficient exact in $x$ (Fig.~\ref{F:zm}) and well as the four-loop one are new.
The contribution of the external flavor ($m=M$) is given by
\begin{eqnarray}
&&\Delta_m(1) = C_F T_F \left(\frac{\alpha_s}{\pi}\right)^2 \biggl[
- \frac{\pi^2-3}{4}
\nonumber\\
&&{} + T_F n_0 \frac{\alpha_s}{\pi}
\left(\zeta_3 + \frac{13}{36} \pi^2 - \frac{11}{12}\right)
\nonumber\\
&&{} - \left(T_F n_0 \frac{\alpha_s}{\pi}\right)^2
\left(\frac{13}{6} \zeta_3 + \frac{4}{45} \pi^4 + \frac{53}{216} \pi^2 - \frac{11}{8}\right)
\nonumber\\
&&{} + \mathcal{O}(\alpha_s^3) \biggr]\,.
\label{2m:Delta1}
\end{eqnarray}
The three- and four-loop terms here agree with~\cite{Marquard:2007uj} and~\cite{Lee:2013sx}.
We do not present lower-loop terms of $z_m$ with positive powers of $\varepsilon$
which may be needed when this ratio is used within calculations containing $1/\varepsilon$ divergences;
these terms can be easily obtained from Eqs.~(\ref{2m:Z}) and~(\ref{2m:B}).

Contributions to $Z_h^{\text{os}}$ with the color structures $C_F T_F^{l-1} n_0^{l-2}$
(i.\,e., the maximum number of quark loops, one of them is massive with mass $m_i$)
can be calculated using Eq.~(\ref{2m:H2}). The results read
\begin{eqnarray}
Z_h^{\text{os}} &=& 1 + C_F \sum_{l=2}^\infty T_F^{l-1} (n_0 P)^{l-2} (l-1) B_h^{(l)}
\nonumber\\
&&{}\times \sum_i \left(\frac{g_0^2 m_i^{-2\varepsilon}}{(4\pi)^{d/2}} \Gamma(\varepsilon)\right)^l
+ \cdots\,,
\nonumber\\
B_h^{(l)} &=&
4 \frac{(3-2\varepsilon) (1+(l-1)\varepsilon) \Gamma^2(1+(l-1)\varepsilon)}%
{l (l-1) (3+2(l-1)\varepsilon) \Gamma(2-\varepsilon) \Gamma^2(1+\varepsilon)}
\nonumber\\
&&{}\times
\frac{\Gamma(1-(l-1)\varepsilon) \Gamma(1+l\varepsilon)}{\Gamma(2+2(l-1)\varepsilon)}\,,
\label{2m:Zh}
\end{eqnarray}
where $g_0 \equiv g_0^{(n_l)}$ and dots denote other color structures.
The ($l=2$)-loop term agrees with~\cite{Broadhurst:1994se},
and the three-loop one with the corresponding color structure in~\cite{Grozin:2006xm}.
According to the regions-based argument in Sect.~\ref{S:Mass},
\begin{equation}
\lim_{x\to0} \left[B_Q^{(l)}(x) - B_{Q0}^{(l)} P - B_h^{(l)} x^{-2l\varepsilon}\right] = 0\,.
\label{2m:BQlim}
\end{equation}

%- }}}
%- {{{ Bibl.:

\bibliography{z}

%- }}}

\end{document}